\documentclass[11pt, oneside]{article} 
\pdfoutput=1

\usepackage{jheppub}
\usepackage{amsmath,amssymb,amsfonts,amsthm}
\usepackage{color}
\usepackage{booktabs}
\definecolor{darkblue}{rgb}{0.1,0.1,.7}
\usepackage{bbm}
\usepackage{enumitem}
\usepackage{graphicx}
\usepackage{latexsym}
\usepackage[labelformat=simple]{subcaption}
\usepackage{amscd}
\usepackage[all,cmtip]{xy}
\usepackage{mathrsfs}
\usepackage{bbold}
\usepackage{slashed}
\usepackage[margin=8pt,font=small,labelfont=bf]{caption}
\usepackage{simplewick}
\usepackage{changepage}
\usepackage[]{algorithm2e}
\usepackage{booktabs,multirow}
\usepackage[OT2,OT1]{fontenc}
\usepackage{braket}
\usepackage{tikz}
\usetikzlibrary{decorations.markings,arrows.meta,decorations.pathmorphing}
\usepackage{pgfplots}
\pgfplotsset{compat=1.10}
\usepgfplotslibrary{fillbetween}
\usetikzlibrary{patterns}

\newcommand*\pFqskip{8mu}
\catcode`,\active
\newcommand*\pFq{\begingroup
	\catcode`\,\active
	\def ,{\mskip\pFqskip\relax}%
	\dopFq
}
\catcode`\,12

\def\dopFq#1#2#3#4#5{%
	\ifnum#1=2\relax
	\ifnum#2=1\relax
	{}_{2}F_{1}\!\left(#3;#4;#5\right)%
	\else
	{}_{#1}F_{#2}\!\left(\genfrac..{0pt}{}{#3}{#4};#5\right)%
	\fi
	\else
	{}_{#1}F_{#2}\!\left(\genfrac..{0pt}{}{#3}{#4};#5\right)%
	\fi
	\endgroup
}
\catcode`,\active

\catcode`\,12

\newcommand{\abs}[1]{\left\lvert#1\right\rvert}

\newcommand{\assign}{:=}

\newcommand{\bb}[2]{ b^{#1}_{#2}}
\newcommand{\hyperF}[4]{{}_2F_1\left(#1,#2;#3;#4\right)}

\newcommand{\RNum}[1]{\uppercase\expandafter{\romannumeral #1\relax}}

\newcommand{\phih}{\hat{\phi}}

\newtheorem{theorem}{Theorem}[section]

\theoremstyle{remark}

\newmuskip\pFqmuskip

\let\a=\alpha 
\let\b=\beta

\let\q=\theta

\let\r=\rho
\let\s=\sigma 
\let\t=\tau  
\let\f=\phi

\let\D=\Delta

\let\F=\Phi

\let\G=\Gamma

\let\del=\partial
\let\<=\langle
\let\>=\rangle

\newcommand{\GG}{\mathcal{G}}

\newcommand{\NN}{\mathcal{N}}
\newcommand{\OO}{\mathcal{O}}
\newcommand{\PP}{\mathcal{P}}

\newcommand{\RR}{\mathcal{R}}

\newcommand{\al}[1]{\begin{align}#1\end{align}}
\newcommand{\spl}[1]{\begin{split}#1\end{split}}
\newcommand{\ga}[1]{\begin{gathered}#1\end{gathered}}
\def \bmat {\begin{pmatrix}}
\def \emat {\end{pmatrix}}

\newcommand{\be}{\begin{equation}}
\newcommand{\ee}{\end{equation}} %added by Xiang for various input shortcuts

\makeatletter
\def\@fpheader{\ }
\makeatother

\title{Locality constraints in AdS$_2$ without parity}
\author[a,b]{Manuel Loparco,}
\author[a]{Gr\'{e}goire Mathys,}
\author[a]{Jo\~ao Penedones,}
\author[c,d]{Jiaxin Qiao,}
\author[a,e]{Xiang Zhao}
\affiliation[a]{Fields and Strings Laboratory, Institute of Physics, \\ 
École Polytechnique Fédéral de Lausanne (EPFL), \\
Route de la Sorge, CH-1015 Lausanne, Switzerland}
\affiliation[b]{Istituto Nazionale di Fisica Nucleare, Sezione di Torino, and\\ 
Department of Physics, 
University of Turin,\\ 
	Via P. Giuria 1, 10125, Turin, Italy}
\affiliation[c]{Laboratory for Theoretical Fundamental Physics, Institute of Physics, \\ 
École Polytechnique Fédérale de Lausanne (EPFL), \\ 
Route de la Sorge, CH-1015 Lausanne, Switzerland}
\affiliation[d]{Kavli Institute for the Physics and Mathematics of the Universe (WPI), \\ 
The University of Tokyo Institutes for Advanced Study, The University of Tokyo, \\ 
Kashiwa, Chiba 277-8583, Japan}
\affiliation[e]{Institut de Physique Théorique, \\ 
Université Paris-Saclay, CEA, CNRS, \\ 
91191, Gif-sur-Yvette, France}

\abstract{
We study bulk locality constraints in quantum field theories in AdS$_2$. 
The  known derivation of locality sum rules in AdS$_{d+1}$ does not apply for $d=1$  due to the different singularity structure of the conformal blocks and the inequivalence of operator orderings on the boundary. Assuming unitarity and a mild growth condition, we establish power-law bounds for correlators, derive dispersion relations and an expansion in terms of ``even'' and ``odd'' local blocks that converges in the entire AdS$_2$. These yield two novel families of symmetric and antisymmetric locality sum rules. 
We test these sum rules explicitly in the free scalar field theory. }

\begin{document}
\maketitle

\section{Introduction}
\label{sec:intro}

Quantum Field Theories (QFTs) in anti–de Sitter (AdS) spacetime provide a powerful theoretical laboratory for exploring fundamental aspects of quantum field theory \cite{Callan:1989em} as well as quantum gravity \cite{Maldacena:1997re,Witten:1998qj,Gubser:1998bc, Aharony:1999ti}. When a theory preserves the full AdS isometry group, the allowed boundary conditions are constrained by conformal symmetry. These constraints render QFTs in AdS far more tractable than their flat-space counterparts. Thus, putting QFTs in AdS facilitates the study of a wide range of interesting phenomena such as phase transitions \cite{Aharony:2012jf,Copetti:2023sya,Ciccone:2024guw,Ciccone:2025dqx} and spontaneous symmetry breaking \cite{Hogervorst:2021spa}, as well as renormalization group (RG) flows \cite{Hogervorst:2021spa,Antunes:2021abs,Lauria:2023uca,Antunes:2024hrt} and constraints imposed by unitarity \cite{Paulos:2016fap} and locality \cite{Levine:2023ywq,Levine:2024wqn,Meineri:2023mps}.

This paper focuses on the recently developed bulk locality constraints~\cite{Levine:2023ywq,Levine:2024wqn,Meineri:2023mps}. The idea behind has previously been applied in the bulk reconstruction program in AdS/CFT (see e.g. \cite{Hamilton:2005ju,Hamilton:2006az,Kabat:2011rz}), as well as in conformal bootstrap \cite{Lauria:2020emq,Bianchi:2022ulu}. Let us briefly review it in the simplest setup for QFT in AdS. Consider a correlator with one bulk operator $\hat\F$ and two boundary operators $\OO_i,\,\OO_j$ in a QFT on AdS$_{d+1}$, i.e. $\braket{\hat{\Phi}(\tau,z)\mathcal{O}_i(\tau_1)\mathcal{O}_j(\tau_2)}$, using standard Poincaré coordinates $z>0$ and $\tau,\tau_1, \tau_2 \in \mathbb{R}^d$. For $d\geqslant2$, $SO^+(1,d+1)$ covariance fixes its form to 
\begin{equation}
\braket{\hat{\Phi}(\tau,z)\mathcal{O}_i(\tau_1)\mathcal{O}_j(\tau_2)}
		=\frac{1}{\abs{\tau_1-\tau_2}^{\Delta_i+\Delta_j}}
		\left(\frac{(\tau-\tau_1)^2+z^2}{(\tau-\tau_2)^2+z^2}\right)^{\tfrac{\Delta_{ji}}{2}}
		\mathcal{G}^{\hat{\Phi}}_{ij}(\chi)\,,
\end{equation}
where $\Delta_{ji} =\D_j-\D_i$ and $\mathcal{G}^{\hat{\Phi}}_{ij}(\chi)$ depends on a single cross-ratio variable $\chi$ (its precise definition will be given shortly).  
A standard analyticity analysis shows that $\mathcal{G}^{\hat{\Phi}}_{ij}(\chi)$ is analytic in the complex domain
\begin{equation*}
		\chi\in\mathbb{C}\backslash(-\infty,0]\,.
\end{equation*}
However, when one expands $\mathcal{G}^{\hat{\Phi}}_{ij}(\chi)$ into conformal blocks ($b^{\hat\F}_l$ and $C_{lij}$ are theory specific data that will be discussed shortly)
\be
\GG_{ij}^{\hat\F}(\chi)=\sum_l b^{\hat\Phi}_l C_{lij} G_{\D_l}^{\D_i\D_j}(\chi)\,,
\ee
each block develops an unphysical singularity near $\chi=1$:
\al{
G_{\D_l}^{\D_i,\D_j}(\chi)\sim
\begin{cases}
    \log(1-\chi)\,,& (d=2)\,, \\
    (1-\chi)^{\frac{2-d}{2}}\,, & (d>2)\,.
\end{cases}
}
Physically, $\chi=1$ is when the bulk operator touches the Euclidean geodesic connecting the two boundary operators and the correlator should be perfectly analytic there. Eliminating these unphysical singularities leads to a new expansion in terms of \emph{local blocks}, and to the corresponding bootstrap equations in AdS, known as the \emph{locality sum rules}~\cite{Levine:2023ywq,Levine:2024wqn,Meineri:2023mps} 
\be
\sum_{l} b^{\hat\Phi}_l C_{lij} \q_n^{(\a)}(\D_l;\D_{ij},\D_{ji};d) = 0\,,\qquad \forall n\in\mathbb{N}_0 \text{ and }\alpha>\frac{\Delta_i+\Delta_j+\Delta_{\hat{\Phi}}}{2}\,,
\label{eq:locality sum rule}
\ee
where $\q_n^{(\a)}$ has a closed form expression and depends on spactime dimension $d$, see \cite[section 3.2]{Levine:2023ywq}.

In this paper we point out that a subtlety arises when $d=1$: \emph{the naive analytic continuation of \eqref{eq:locality sum rule} to $d=1$ does not always hold}. The issue is closely analogous to the following toy example. Suppose we have a rotation-invariant smooth function on $\mathbb{R}^d$, denoted by $f(x)$.  
For $d\geqslant2$, rotation invariance implies that $f(x)$ depends only on $\abs{x}$, and smoothness then implies that $f(x)$ depends only on $y=x^2$.  
It follows that the Taylor expansion of $f(x)$ must take the form
\begin{equation}
	\begin{split}
		f(x)=\sum_{n=0}^{N}a_n y^n+O\left(y^{N+1}\right),\qquad (y=x^2)\,.
	\end{split}
\end{equation}
However, if we consider the special case $d=1$, rotation invariance plays no role, and $f(x)$ simply admits a Taylor expansion of the form
\begin{equation}
	\begin{split}
		f(x)=\sum_{n=0}^{N}b_n x^n+O\left(x^{N+1}\right)\,.
	\end{split}
\end{equation}
If we rewrite it in terms of $y$, we find
\begin{equation}
	\begin{split}
		f(x)=\sum_{n=0}^{N}b_{2n} y^n+\text{sgn}(x)\sum_{n=0}^{N}b_{2n+1} y^{n+1/2}+O\left(x^{2N+2}\right)\,.
	\end{split}
\label{eq:y expansion of f(x)}
\end{equation}
We see that a square-root singularity in $y$ appears in the expansion, although nothing is actually singular: this is merely a coordinate artifact, and the function remains perfectly smooth in $x$.

Similar to the toy example above, $1-\chi$ plays the role of $y$ in AdS$_2$, while $\sqrt{1-\chi}$ mimics $x$. A physical correlator in general admits a Taylor expansion in $\sqrt{1-\chi}$, and so does each block itself. Therefore, superficially, there is nothing unphysical to remove.\footnote{ If we further impose parity, which is analogous to considering $f(x)$ to be an even or odd function of $x$ in the toy example above, then the Taylor expansion of the block and of the correlator in general do not have the same structure. See section \ref{subsec:parity} for more details.} Nevertheless, the standard conformal block expansion in AdS$_2$ still fails to converge at $\chi=1$, where the physical correlator should remain regular. 

This is closely related to the fact that the operator orderings on the 1d conformal boundary are, in general, equivalent only up to cyclic permutations. In particular, for the correlator discussed above its block decomposition involves either $C_{lij}$ or $C_{lji}$($\neq C_{lij}$ in general), depending on which side of the geodesic $\chi=1$ the bulk operator is at. The abrupt change of the OPE coefficient to use in the conformal block expansion is another imprint of the divergence of conformal block expansion at $\chi=1$. The construction of a new block expansion that can be term-by-term analytically continued smoothly across $\chi=1$, sharing the same analyticity of the correlator itself, once again leads to locality sum rules. This constitutes the main focus of the present work. 

The main result is the following new sum rules for QFTs in AdS$_2$
\begin{align}
		&\sum_{l}b^{\hat{\Phi}}_{l}\,
		\frac{C_{ijl}+C_{jil}}{2}\theta^{(\alpha)}_n(\Delta_l;\Delta_{ij},\Delta_{ji})=0,\qquad& \alpha&>\frac{1}{2}(\Delta_i+\Delta_j+\Delta_{\hat{\Phi}})\,,
\\
		&\sum_{l}b^{\hat{\Phi}}_{l}\,
		\frac{C_{ijl}-C_{jil}}{2}\theta^{(\alpha)}_n(\Delta_l;\Delta_{ij}+1,\Delta_{ji}+1)=0, \qquad &\alpha&>\frac{1}{2}(\Delta_i+\Delta_j+\Delta_{\hat{\Phi}}-1)\,,
\end{align}
for any non-negative integer $n$ and $\q_n^{(\a)}$ is given in \eqref{def:theta}. These sum rules do not assume parity invariance.

The results presented in this paper are foundational to our companion work \cite{Loparco:2026fki}, which studies  RG flows  of QFTs in AdS$_2$. In that context, integrated correlators are required, but the integration of blocks over AdS$_2$ does not commute with the standard conformal block expansion. The local-block expansion and the associated sum rules developed in this paper will play a crucial role in resolving this issue and allowing for the exchange of sums and integrals. 

We also note that a closely related issue arises in the context of conformal defects. For the massless free scalar field in $d>2$ and codimension $q\geqslant2$, it was shown in \cite{Lauria:2020emq} that almost all conformal defects are trivial in the sense that they are described by generalized free theories, except for a few special cases. In the case of conformal line defects, the argument of \cite{Lauria:2020emq} assumes parity symmetry. The recent paper \cite{Bartlett-Tisdall:2025iqx} pointed out that if one relaxes the parity invariance, then the free scalar field, in fact, can admit nontrivial conformal line defects. The reason why the line defect is special compared to higher-dimensional defects is exactly the same as in the AdS$_2$ case discussed above: the existence of the cross ratio $\sqrt{1-\chi}$, which is smooth for a one-dimensional defect but not in higher dimensions. The fact that parity invariance can lead to a stronger constraint is explained in section \ref{subsec:parity}.

\paragraph{Structure of the paper} In section \ref{sec:Bkbdbd correlator}, we explain the setup and combine unitarity with both radial quantization and equal-time quantization to show a uniform power-law bound of the bulk-boundary-boundary correlator. In section \ref{sec:locality constraints}, we derive a dispersion relation for the correlator, construct new local blocks and derive a new set of locality sum rules. In section \ref{sec:test}, we test the sum rules using free scalar theory in AdS$_2$. Conclusion is given in section \ref{sec:conclusion}. In appendix \ref{app:PL}, we give a brief introduction to the sector version of the Phragmén–Lindelöf theorem, which is used to prove the power-law bound.

\section{\texorpdfstring{Bulk-boundary-boundary correlator in AdS$_2$}{Bulk-boundary-boundary correlator in AdS2}}
\label{sec:Bkbdbd correlator}

\subsection{Geometry}
Let us consider two-dimensional hyperbolic space, or Euclidean AdS$_2$, in local Poincar\'e coordinates. The metric is
% It will be useful to introduce the local system of coordinates known as the Poincar\'e patch
% \begin{equation}
% \begin{aligned}
%     X^0&=\frac{1+z^2+\tau^2}{2z}\,,&  \qquad X^1&=\frac{\tau}{z}\,,&\qquad X^2&=\frac{1-z^2-\tau^2}{2z}\,,\\
%     P^0&=\frac{1+\tau^2}{2}\,,\qquad & P^1&=\tau\,,&\qquad  P^2&=\frac{1-\tau^2}{2}\,.
% \end{aligned}
% \end{equation}
\begin{equation}
	\begin{split}
	 ds^2=\frac{d\tau^2+dz^2}{z^2}\,,\qquad\qquad  z>0\,.\label{eq:AdSPoicare}
	\end{split}
\end{equation}
We always set the AdS radius  to 1. 
% The Euclidean AdS$_2$ can be embedded into the three-dimensional Lorentzian space $\mathbb{R}^{1,2}$ via the constraint
% \begin{equation}
%     -(X^0)^2+(X^1)^2+(X^2)^2=-1\,.
% \end{equation}
% We conventionally choose the branch with $X^0>0$. Then the explicit embedding in terms of Poincaré corrdinates is given by
% \begin{equation}
%     (X^0,X^1,X^2)=\left(\frac{1+z^2+\tau^2}{2 z},\frac{\t}{z},\frac{1-z^2-\tau^2}{2 z}\right)\in\mathbb{R}^{1,2}\,.
% \end{equation}

The isometry group of AdS$_2$ is $O^+(1,2)$. It contains two disconnected pieces
\begin{equation}
	\begin{split}
		O^+(1,2)=SO^+(1,2)\sqcup \mathcal{P}SO^+(1,2),
	\end{split}
\end{equation}
with $\mathcal{P}$ the usual parity transformation
\begin{equation}\label{def:parity}
	\begin{split}
		\mathcal{P}:(\t,z) \mapsto  (-\t, z)\,.
	\end{split}
\end{equation}
The connected subgroup $SO^+(1,2)$ is isomorphic to $PSL(2;\mathbb{R})$, which is realized by
\begin{equation}
	\begin{split}
		w=\tau+i z
		\quad\mapsto\quad w'=\frac{aw+b}{cw+d},\qquad\text{with}\qquad \left(\begin{matrix}
			a & b \\
			c & d \\
		\end{matrix}\right)\in SL(2;\mathbb{R}),
	\end{split}\label{eq:Sl2action}
\end{equation}
such that $ad-bc=1$. After an $SL(2;\mathbb{R})$ transformation, the coordinate $z'$ remains positive
\begin{equation}
	z' = \frac{(a d - b c) z}{c^2 z^2+(c \tau +d)^2} = \frac{z}{c^2 z^2+(c \tau +d)^2}\, .
\end{equation}
We will also need the inversion
\al{
\RR:
(\t,z)\mapsto (\t',z')=\left(\frac{\t}{\t^2+z^2},\frac{z}{\t^2+z^2}\right)\,.
\label{eq:geo inversion}}
Notice that $\PP$ and $\RR$ belong to the $\PP SO^{+}(1,2)$ part, rather than $SO^{+}(1,2)\cong SL(2,\mathbb{R})$ part, of the AdS$_2$ isometry group, and one can easily check that
\al{
\PP: ds^2\to ds^2\,,\qquad\quad
\RR: ds^2 \mapsto \frac{d\t'^2+dz'^2}{z'^2} = \frac{d\t^2+dz^2}{z^2}=ds^2\,.
}
\subsubsection{Embedding space}
It is also useful to consider the embedding of AdS$_2$ in three-dimensional Minkowski space. In this case, the points in the embedding space are indicated by $X^A=(X^0,X^1,X^2)\in \mathbb{R}^{1,2}$ and belong to the hyperboloid
\begin{equation}
    -(X^0)^2+(X^1)^2+(X^2)^2=-1\,, \qquad X^0>0\, .
\end{equation}
We conventionally choose the upper branch with $X^0>0$ of the hyperboloid. Then the explicit embedding in terms of Poincaré coordinates is given by
 \begin{equation}
     (X^0,X^1,X^2)=\left(\frac{1+z^2+\tau^2}{2 z},\frac{\t}{z},\frac{1-z^2-\tau^2}{2 z}\right)\in\mathbb{R}^{1,2}\,.
 \end{equation}
Also, points on the conformal boundary are embedding space lightrays, denoted as $P^A=(P^0,P^1,P^2)\in\mathbb{R}^{1,2}$, which  satisfy 
\begin{equation}
    -(P^0)^2+(P^1)^2+(P^2)^2=0\,, \qquad P^0>0\,,
\end{equation}
with the identification $P^A\sim\lambda P^A$. They are given in terms of Poincar\'e coordinates as 
 \begin{equation}
    (P^0,P^1,P^2) = \left(\frac{1+\tau^2}{2}\,,\tau\,, \frac{1-\tau^2}{2}\right) \in \mathbb{R}^{1,2}\, .
 \end{equation}
Now the actions of $\mathcal{P}$ and $\mathcal{R}$ become
\be
\PP: (X^0,X^1,X^2) \mapsto (X^0,-X^1,X^2)\,,
\qquad
\RR: (X^0,X^1,X^2) \mapsto (X^0,X^1,-X^2)\,,
\ee
and similarly for the boundary points $P$. They are obviously related by a rotation in the $(X^1,X^2)$ plane.

This construction is particularly useful as the group of isometries that preserve one branch of the hyperboloid ($O^+(1,2)$) is realized linearly in embedding space. Consequently, all invariant cross-ratios can be constructed through dot products of the $X_i$ and $P_j$ vectors with the embedding space metric $\eta_{AB}=\text{diag}(-1,1,1)$. 

In terms of the embedding space, it is immediately clear that any three points allow an $SO^+(1,2)$ invariant via the Levi-Civita tensor:
\begin{equation}
    \epsilon(X_1,X_2,X_3),\ \frac{\epsilon(X_1,P_2,P_3)}{P_2\cdot P_3},\ldots\,.
\end{equation}
The invariants constructed in this way are not $O^+(1,2)$ invariant because the Levi-Civita tensor changes sign under parity transformation. In contrast, in higher dimensions, any $SO^+(1,d)$ invariant constructed from three points is manifestly an $O^+(1,d)$ invariant.

\subsection{Conformal block expansion}

 Using $SO^+(1,2)$ invariance, the bulk-boundary-boundary correlator $\braket{\hat{\Phi}(\tau,z)\mathcal{O}_i(\tau_1)\mathcal{O}_j(\tau_2)}$, where we denote a bulk field as $\hat{\Phi}(\tau,z)$, can be fixed to
\begin{equation}\label{bkbdbd:generalform}
	\braket{\hat{\Phi}(\tau,z)\mathcal{O}_i(\tau_1)\mathcal{O}_j(\tau_2)}
	=\frac{1}{\abs{\tau_1-\tau_2}^{\Delta_i+\Delta_j}}
	\left(\frac{(\tau-\tau_1)^2+z^2}{(\tau-\tau_2)^2+z^2}\right)^{\tfrac{\Delta_{ji}}{2}}
	\mathcal{G}^{\hat{\Phi}}_{ij}(\rho)\,,
\end{equation}
where $\Delta_{ji}:=\Delta_j-\Delta_i$, and $\rho$ is an $SO^+(1,2)$-invariant defined by 
\begin{equation}\label{def:rho}
	\begin{split}
		\rho&:=\frac{2\,\epsilon(X,P_1,P_2)}{\sqrt{(-2X\cdot P_1)(-2X\cdot P_2)(-2P_1\cdot P_2)}} \\
        &=\text{sgn}(\tau_1-\tau_2)\frac{(\tau-\tau_1)(\tau-\tau_2)+z^2}{\sqrt{\left((\tau-\tau_1)^2+z^2\right)\left((\tau-\tau_2)^2+z^2\right)}}\,,
	\end{split}
\end{equation}
with $\epsilon$ the Levi-Civita tensor in the three-dimensional embedding space with convention $\epsilon_{012}=1$. In the special configuration $\tau_1=0$ and $\tau_2=\infty$, this simplifies to $\rho=\tfrac{\tau}{\sqrt{\tau^2+z^2}}$.

\begin{figure}[t]
	\centering
	\begin{tikzpicture}
			% Draw axes
			\draw[->,thick] (-4,0) -- (4,0) node[right] {$\tau$};
			
			% Draw semicircle from (-2,0) to (2,0)
			\draw[thick,dashed] (-2,0) arc[start angle=180, end angle=0, radius=2];
			
			% Add two points
			\fill (-2,0) circle (2pt) node[below] {$\mathcal{O}_i$};
			\fill (2,0) circle (2pt) node[below] {$\mathcal{O}_j$};

            % Add values of rho
			\node at (0,1) {$0<\rho<1$};
            \node at (-3,1) {$-1<\rho<0$};
            \draw[->,thick] (1.5,2.2) -- (1.2,1.7) ;
            \node at (1.6,2.4) {$\rho=0$};
            \draw[->,thick] (-3,-0.5) -- (-3,-0.05) ;
            \node at (-3,-0.7) {$\rho=-1$};
            \draw[->,thick] (0,-0.5) -- (0,-0.05) ;
            \node at (0,-0.7) {$\rho=1$};
            \draw[->,thick] (3,-0.5) -- (3,-0.05) ;
            \node at (3,-0.7) {$\rho=-1$};
		\end{tikzpicture}
	\caption{The range of $\rho$ depending on the position of the bulk point. Here the two boundary points are fixed. Exchanging the positions of $\mathcal{O}_i$ and $\mathcal{O}_j$ leads to $\rho\rightarrow-\rho$.}
	\label{fig:rho}
\end{figure}
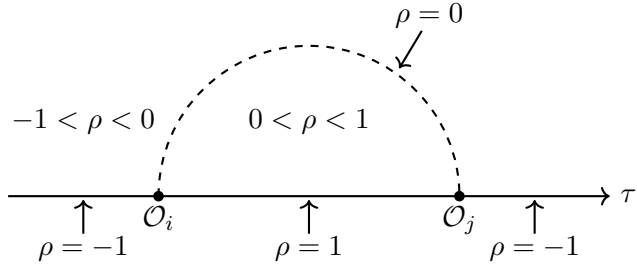

Generically, a bulk operator $\hat{\Phi}(\tau,z)$ can be expanded in terms of boundary operators $\mathcal{O}_i(\tau)$ with finite radius of convergence, and  this is called the \emph{boundary operator expansion} (BOE). It takes the form \cite{Paulos:2016fap,Levine:2023ywq}
\begin{equation}\label{eq:BOE}
	\hat{\Phi}(\tau,z)
	=\sum\limits_{i}\bb{\hat{\F}}{i}\sum\limits_{n=0}^{\infty}
	\frac{(-1)^n}{n!\,4^n\,(\Delta_i+\tfrac{1}{2})_{n}}\,z^{\Delta_i+2n}\,
	\partial_\tau^{2n}\mathcal{O}_i(\tau)\,, 
\end{equation}
where the index $i$ runs over boundary primaries $\OO_i$, while the sum over $n$ incorporates the contributions of their descendants. The coefficients $b^{\hat{\F}}_{i}$ are called BOE coefficients, and they appear in the bulk-boundary two-point functions
\begin{equation}\label{def:bkbdtwopt}
	\braket{\hat{\F}(\tau_1,z_1)\mathcal{O}_i(\tau_2)}
	=\bb{\hat{\F}}{i}\left(\frac{z_1}{(\tau_1-\tau_2)^2+z_1^2}\right)^{\Delta_i}\,.
\end{equation}

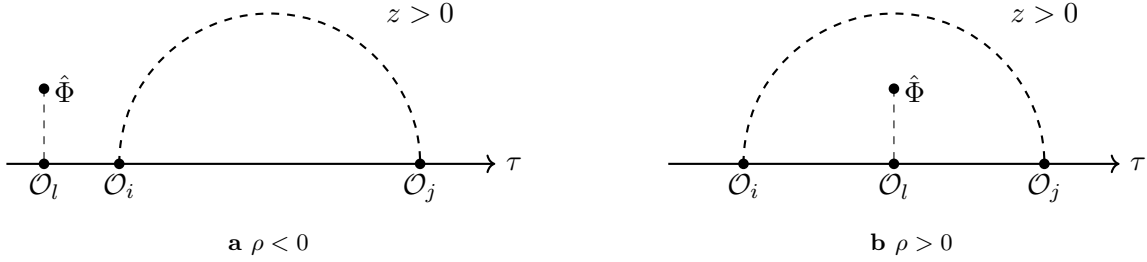
\begin{figure}[t]
	\centering
	\begin{subfigure}[b]{0.45\textwidth}
		\centering
		\begin{tikzpicture}
			% Draw axes
			\draw[->,thick] (-3.5,0) -- (3,0) node[right] {$\tau$};
			
			% Draw semicircle from (-2,0) to (2,0)
			\draw[thick,dashed] (-2,0) arc[start angle=180, end angle=0, radius=2];
			
			% Add label for the region
			\node at (2,2) {$z > 0$};
			
			% Add two points
			\fill (-2,0) circle (2pt) node[below] {$\mathcal{O}_i$};
			\fill (2,0) circle (2pt) node[below] {$\mathcal{O}_j$};
			\fill (-3,0) circle (2pt) node[below] {$\mathcal{O}_l$};
			
			% Add point at (-3,1)
			\fill (-3,1) circle (2pt) node[right] {$\hat{\Phi}$};
			
			% Add dashed line from (-3,1) to tau axis
			\draw[dashed] (-3,1) -- (-3,0);
		\end{tikzpicture}
		\caption{$\rho<0$}
		\label{fig:bkbdbd1}
	\end{subfigure}
	\hfill
	\begin{subfigure}[b]{0.45\textwidth}
		\centering
		\begin{tikzpicture}
			% Draw axes
			\draw[->,thick] (-3,0) -- (3,0) node[right] {$\tau$};
			
			% Draw semicircle from (-2,0) to (2,0)
			\draw[thick,dashed] (-2,0) arc[start angle=180, end angle=0, radius=2];
			
			% Add label for the region
			\node at (2,2) {$z > 0$};
			
			% Add two points
			\fill (-2,0) circle (2pt) node[below] {$\mathcal{O}_i$};
			\fill (2,0) circle (2pt) node[below] {$\mathcal{O}_j$};
			\fill (0,0) circle (2pt) node[below] {$\mathcal{O}_l$};
			
			% Add point at (0,1)
			\fill (0,1) circle (2pt) node[right] {$\hat{\Phi}$};
			
			% Add dashed line from (0,1) to tau axis
			\draw[dashed] (0,1) -- (0,0);
		\end{tikzpicture}
		\caption{$\rho>0$}
		\label{fig:bkbdbd2}
	\end{subfigure}
	\caption{Conformal block expansion for different types of configurations. The choice of $C_{ijl}$ vs $C_{jil}$ depends on the cyclic ordering of the boundary operators. For figure \ref{fig:bkbdbd1}, the cyclic ordering is $[lij]=[ijl]=[jli]$, so the corresponding OPE coefficient is $C_{ijl}$. While for figure \ref{fig:bkbdbd2}, the corresponding OPE coefficient is $C_{jil}$.}
	\label{fig:bkbdbdconfigs}
\end{figure}

The $SO^+(1,2)$-invariant function $\mathcal{G}^{\hat{\Phi}}_{ij}(\rho)$ can be further decomposed into conformal blocks using the BOE \eqref{eq:BOE}. Depending on the sign of $\rho$, the expansion takes the form\footnote{In this paper, we use the convention that $\braket{\mathcal{O}_i(0)\mathcal{O}_j(1)\mathcal{O}_k(\infty)}=C_{ijk}$ in the boundary CFT, where $\mathcal{O}_k(\infty):=\lim\limits_{\tau\rightarrow\infty}\left[|\tau|^{2\Delta_k}\mathcal{O}_k(\tau)\right]$. Note that in 1D CFT, $C_{ijk}$ is symmetric under cyclic permutation of the indices, but it is not necessarily the same as $C_{jik}$. If we impose parity symmetry, then $C_{ijk}=\pm C_{jik}$, see section \ref{subsec:parity}.}
\begin{equation}\label{bkbdbd:exp}
	\mathcal{G}^{\hat{\Phi}}_{ij}(\rho)=
	\begin{cases}
		\sum\limits_{l}\bb{\hat{\Phi}}{l}C_{ijl}
		G_{\D_l}^{\D_i\D_j}(\chi)\,, & \quad \text{for }\rho<0\,, \\[6pt]
		\sum\limits_{l}\bb{\hat{\Phi}}{l}C_{jil}
		G_{\D_l}^{\D_i\D_j}(\chi)\,, & \quad \text{for }\rho>0\,,
	\end{cases}
\end{equation}
where the reasoning for the different expansions involving $C_{ijl}$ or $C_{jil}$ is explained in figure \ref{fig:bkbdbdconfigs}. The cross ratio $\chi$ appearing in \eqref{bkbdbd:exp} is the $O^+(1,2)$ invariant defined by 
\begin{equation}\label{def:chi}
	\chi
	:=\frac{-2 P_1\cdot P_2}{(-2 P_1\cdot X)(-2 P_2\cdot X)}=\frac{(\tau_{1}-\tau_{2})^2z^2}
	{[(\tau-\tau_1)^2+z^2][(\tau-\tau_2)^2+z^2]}\,.
\end{equation}
Note that in Euclidean AdS, $\chi\in[0,1]$ and $\r\in[-1,1]$, and that the two cross-ratios $\rho$ and $\chi$ are related as 
\begin{equation} 
\chi=1-\rho^2\, .\label{eq:chifromrho}
\end{equation}
The conformal block $G_{\D_l}^{\D_i\D_j}(\chi)$ in \eqref{bkbdbd:exp} is given by
\begin{equation}\label{def:bkbdbd block}
	G_{\D_l}^{\D_i\D_j}(\chi)
	=\chi^{\tfrac{\Delta_l}{2}}
	\hyperF{\tfrac{\Delta_{lij}}{2}}{\tfrac{\Delta_{lji}}{2}}
	{\Delta_l+\tfrac{1}{2}}{\chi}\,,
\end{equation}
where $\D_{ijk}\equiv\D_i+\D_j-\D_k$.  
The form of the conformal block $G_{\D_l}^{\D_i\D_j}(\chi)$ can be obtained either by explicitly summing over descendants in the OPE \cite{Lauria:2020emq} or by solving the quadratic Casimir equation \cite{Meineri:2023mps}. 

Note that neither of the expansions in \eqref{bkbdbd:exp} converges at $\rho=0$. In this configuration, the bulk point and the two boundary points lie on the same geodesic in AdS$_2$, even though nothing singular occurs physically. When the expansion \eqref{bkbdbd:exp} is organized in increasing order of the exchanged scaling dimensions $\Delta_l$, it converges exponentially fast for $\chi<1$ but diverges as a power law at $\chi=1$.\footnote{For example, in the free scalar theory, one can explicitly verify that
\begin{equation*}
    \abs{b^{\hat{\Phi}}_l}\sim \Delta_l^{\Delta_{\hat{\Phi}}-3/4},\qquad 
	\abs{C_{ijl}}\sim 2^{-\Delta_l}\Delta_l^{\Delta_i+\Delta_j-3/4} 
	\quad\text{as }\Delta_l\rightarrow\infty\,.
\end{equation*}
Here, the factors that are independent of $\Delta_l$ are omitted. In fact, these asymptotic behaviors hold in an averaged sense in any unitary theory. Using the large-$\Delta_l$ behavior of the conformal block at $\chi=1$,
\begin{equation*}
    G_{\Delta_l}^{\Delta_i\Delta_j}(1)=\frac{\sqrt{\pi } \Gamma \left(\Delta_l+\frac{1}{2}\right)}{\Gamma \left(\tfrac{\Delta_{lij}+1}{2}
   \right) \Gamma \left(\tfrac{\Delta_{lji}+1}{2}\right)}\sim 2^{\Delta_l}\,,
\end{equation*}
we obtain
\begin{equation*}
    \sum\limits_{\Delta_l\leqslant\Delta_*}\abs{b^{\hat{\Phi}}_l\,C_{ijl}G_{\Delta_l}^{\Delta_i\Delta_j}(1)}\sim\sum\limits_{\Delta_l\leqslant\Delta_*}\Delta_l^{\Delta_i+\Delta_j+\Delta_{\hat{\Phi}}-3/2}\, ,
\end{equation*}
which diverges as a power law when $\Delta_i+\Delta_j+\Delta_{\hat{\Phi}}>\tfrac{1}{2}$.}
% \JP{as we will show in section \ref{sec:section231}.}
% \GM{Maybe not what you had in mind? }\JQ{I added the footnote for this point.}

\subsection{Quantization and analyticity}
The generators of AdS$_2$ isometries are the dilatation $D$, translation $P$ and special conformal transformation $K$. They satisfy the $\mathfrak{sl}(2,\mathbb{R})$ algebra
\al{
[D,P]= P\,,\qquad [D,K]=-K\,,\qquad [K,P]=2D\,.
}
In Poincar\'{e} coordinates they act on local scalar operators as differential operators
\al{\spl{
D=-(\tau\partial_\tau+z\partial_z)\,,\qquad
P=\del_\t\,,\qquad
K=(\t^2-z^2)\del_\t + 2\t z \del_z \,.
}}

In AdS$_2$, it is often convenient to use the radial quantization picture, which foliates the upper half plane into semicircles, with the Hamiltonian given by the dilatation generator $D$. 
In this framework, energy corresponds to the scaling dimension.

Alternatively, one may consider the equal-time quantization, which foliates AdS$_2$ into constant-$\tau$ slices and takes the time-translation generator $P$ as the Hamiltonian.

\begin{figure}[!t]
    \centering
    \includegraphics[width=0.85\linewidth]{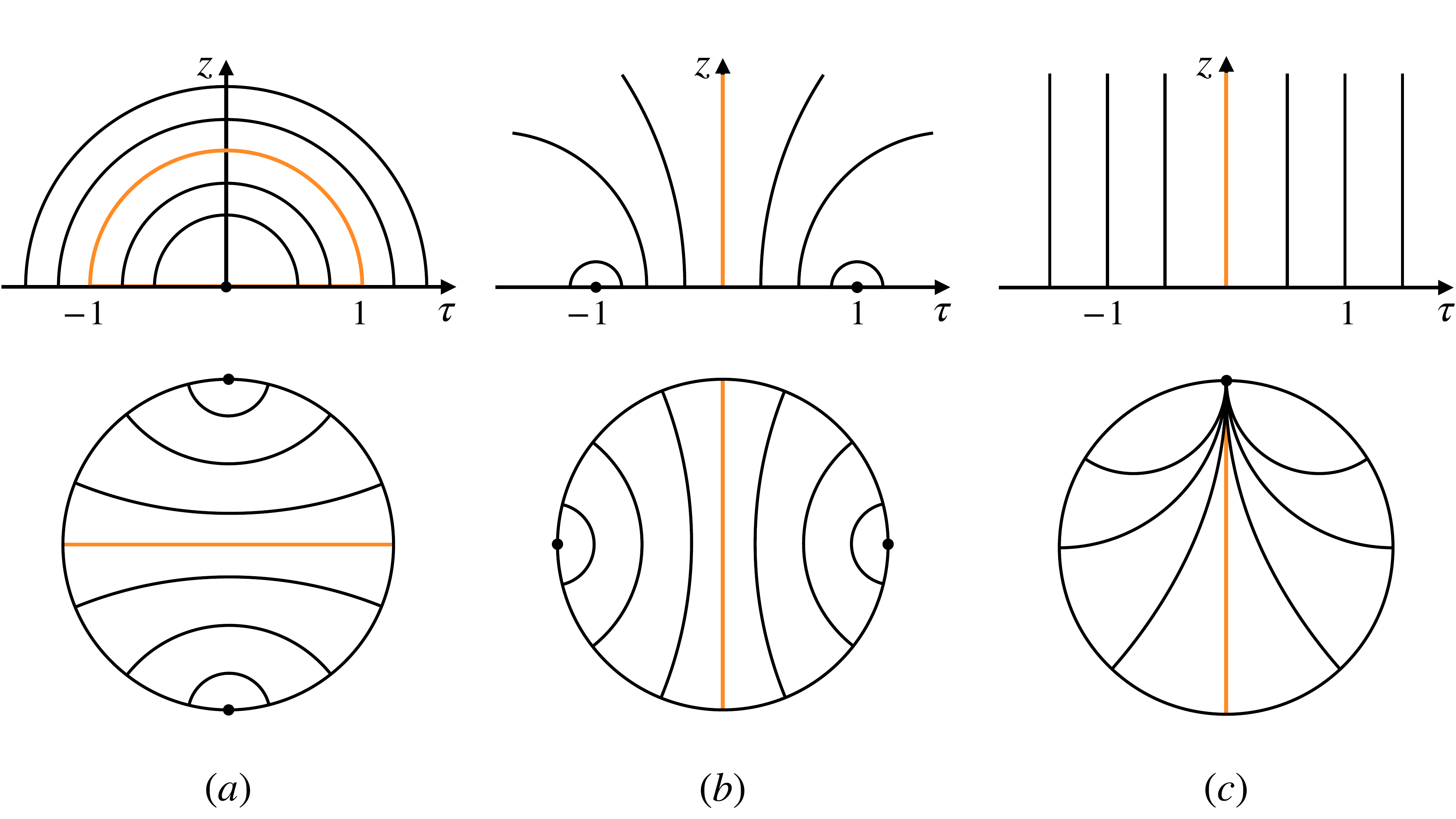}
    \caption{Three types of quantizations in the upper half plane (upper row) and in the Poincar\'{e} disk (lower row): $(a)$ Radial quantization; $(b)$ North-South quantization; $(c)$ Equal-time quantization.}
    \label{fig:quantization}
\end{figure}

These two quantization schemes are not conformally equivalent: the spectrum of $D$ is typically discrete, whereas that of $P$ is continuous. Nevertheless, their unitarity conditions coincide. This equivalence follows from the fact that radial quantization is conformally related to the North–South (N–S) quantization \cite{Luscher:1974ez} (see also \cite[section 3.1.5]{Rychkov:2016iqz}), so their unitarity conditions are equivalent. Furthermore, the unitarity condition in the N–S quantization is the same as that of the equal-time quantization. See figure \ref{fig:quantization}.

The existence of a Hilbert space and the unitarity condition play a crucial role in determining the analytic properties of correlators~\cite{Osterwalder:1973dx,Osterwalder:1974tc,Glaser:1974hy,Luscher:1974ez,Pappadopulo:2012jk} and of scattering amplitudes \cite{Eden:1971fm,Eden:1966dnq}. 
In what follows, we establish the analytic properties of the correlator from the viewpoint of both quantization pictures. 
As we shall see, the analyticity domains obtained from the two quantization schemes are different, and each provides complementary insight.

Then, we will make a very mild assumption about the growth of the correlator as it approaches the light-cone singularity. With this assumption, we will derive a uniform upper bound, which is important for establishing the local-block expansion and the locality sum rules in the next sections.

\subsubsection{Analyticity from radial quantization \label{sec:section231}}

In radial quatization, the Hermitian conjugates of the generators are
\al{
D^\dagger = D\,,\qquad P^\dagger = K\,,\qquad K^\dagger = P\,,
}
and the Hermitian conjugates of operators are
\al{\ga{
\hat\Phi(\t,z)^\dagger = \hat\Phi\left(\RR(\t, z)\right)\,,
\qquad
\OO_i(\t)^\dagger = |\t|^{-2\D_i}\OO(\RR\t)\,,
}}
where the inversion $\RR$ is given in \eqref{eq:geo inversion}.

To understand the analytic properties of the correlator, it is useful to introduce a new variable
\begin{equation}
	\begin{split}
		\xi:=\sqrt{\frac{1-\rho}{1+\rho}}\,.
	\end{split}
\end{equation}
Geometrically, $\xi$ corresponds to a conformal mapping of the generic configuration in \eqref{bkbdbd:generalform} into the following one \cite{Bianchi:2022ulu,Levine:2023ywq}\footnote{The variable $\xi$ used in this paper corresponds to $1/\rho$ in \cite{Levine:2023ywq} and to the variable $r$ in \cite{Bianchi:2022ulu}.}
\begin{equation}\label{config:special}
	\begin{split}
		(\tau,z)=(0,\xi),\qquad\tau_1=-1,\qquad\tau_2=1\,.
	\end{split}
\end{equation}

In the radial quantization picture, a bulk–boundary–boundary correlator admits a natural series expansion in $\xi$. When $\xi<1$, we can expand the correlator as 
\begin{equation}
	\begin{split}
		\braket{\hat{\Phi}(0,\xi)\mathcal{O}_i(-1)\mathcal{O}_j(1)}&=\sum\limits_{l}\sum\limits_{n=0}^{\infty}\braket{\mathcal{O}_i(-1)\mathcal{O}_j(1)|l,n}\braket{l,n|\hat{\Phi}(0,\xi)} \\
		&=\sum\limits_{l}\sum\limits_{n=0}^{\infty}\braket{\mathcal{O}_i(-1)\mathcal{O}_j(1)|l,n}\braket{l,n|\hat{\Phi}(0,1)}\xi^{\Delta_l+n}.
	\end{split}
\label{eq:correlator as a series}
\end{equation}
In this expansion, $l$ labels the primary states and $n$ the descendants of a given primary. This expansion is absolutely convergent for $\abs{\xi}<1$. To see this, we use the unitarity condition of the theory, which implies the Cauchy–Schwarz inequality: 
% \begin{equation}
% 	\begin{split}
% 		\braket{\hat{\Phi}(0,\xi)\mathcal{O}_i(-1)\mathcal{O}_j(1)}^2&=\left(\sum\limits_{l,n}\abs{\braket{\mathcal{O}_i(-1)\mathcal{O}_j(1)|l,n}\braket{l,n|\hat{\Phi}(0,1)}\xi^{\Delta_l+n}}\right)^2 \\
% 		&\leqslant \left(\sum\limits_{l,n}\abs{\braket{\mathcal{O}_i(-1)\mathcal{O}_j(1)|l,n}}^2\abs{\xi}^{\Delta_l+n}\right)  \times\left(\sum\limits_{l,n}\abs{\braket{l,n|\hat{\Phi}(0,1)}}^2\abs{\xi}^{\Delta_l+n}\right) \\
% 		&=\braket{\mathcal{O}_i(-1)\mathcal{O}_i(-\abs{\xi})\mathcal{O}_j(\abs{\xi})\mathcal{O}_j(1)}\abs{\xi}^{\Delta_i+\Delta_j}\times\braket{\hat{\Phi}(0,\abs{\xi})\hat{\Phi}(0,1)} \\
% 		&\leqslant\frac{C}{(1-\abs{\xi})^{2(\Delta_i+\Delta_j+\Delta_{\hat{\Phi}})}},
% 	\end{split}
% \end{equation}
% \begin{equation}
% \begin{split}
% \left|\braket{\hat{\Phi}(0,\xi)\mathcal{O}_i(-1)\mathcal{O}_j(1)}\right|^2
% &=
% \abs{\sum\limits_{l,n}\braket{\mathcal{O}_i(-1)\mathcal{O}_j(1)|l,n}\braket{l,n|\hat{\Phi}(0,1)}\xi^{\Delta_l+n}}^2 \\
% &\leqslant \left(\sum\limits_{l,n}\abs{\braket{\mathcal{O}_i(-1)\mathcal{O}_j(1)|l,n}}^2\abs{\xi}^{\Delta_l+n}\right)  \times\left(\sum\limits_{l,n}\abs{\braket{l,n|\hat{\Phi}(0,1)}}^2\abs{\xi}^{\Delta_l+n}\right) \\
% &=
% \braket{\mathcal{O}_i(-1)\mathcal{O}_i(-\abs{\xi})\mathcal{O}_j(\abs{\xi})\mathcal{O}_j(1)}\abs{\xi}^{\Delta_i+\Delta_j}\times\braket{\hat{\Phi}(0,\abs{\xi})\hat{\Phi}(0,1)} \\
% &\leqslant
% \frac{C}{(1-\abs{\xi})^{2(\Delta_i+\Delta_j+\Delta_{\hat{\Phi}})}},
% \end{split}
% \end{equation}
\begin{equation}
\begin{split}
\left|\braket{\hat{\Phi}(0,\xi)\mathcal{O}_i(-1)\mathcal{O}_j(1)}\right|^2
&=
\abs{ \braket{\mathcal{O}_i(-1)\mathcal{O}_j(1) \xi^{D}\hat{\Phi}(0,1)}}^2=
\abs{ \braket{\mathcal{O}_i(-1)\mathcal{O}_j(1) \xi^{D/2} |\xi^{D/2}\hat{\Phi}(0,1)}}^2\\
&\leqslant 
\braket{\mathcal{O}_i(-1)\mathcal{O}_j(1) \abs{\xi}^{D}\mathcal{O}_j(1)\mathcal{O}_i(-1) }\times
 \braket{\hat{\Phi}(0,1)\abs{\xi}^{D}\hat{\Phi}(0,1)}
 \\
&=
\braket{\mathcal{O}_i(-1)\mathcal{O}_i(-\abs{\xi})\mathcal{O}_j(\abs{\xi})\mathcal{O}_j(1)}\abs{\xi}^{\Delta_i+\Delta_j}\times\braket{\hat{\Phi}(0,\abs{\xi})\hat{\Phi}(0,1)} \\
&\leqslant
\frac{C}{(1-\abs{\xi})^{2(\Delta_i+\Delta_j+\Delta_{\hat{\Phi}})}},
\end{split}
\end{equation}
% \XZ{the second equality follows from $\abs{\sum_{n} u^*_n v_n}^2 \leq \sum_n \abs{u_n}^2 \sum_n \abs{v_n}^2$, here we choose $u_n^*=\<\OO_i(-1)\OO_j(1)|l,n\>|\xi|^{(\D_l+n)/2}$ and $v_n=\<l,n|\hat\F(0,1)\>\abs{\xi}^{(\D_l+n)/2}$, but Joao's version is more compact without writing the basis for $\vec{u}$ and $\vec{v}$ explicitly.}
where $\Delta_{\hat{\Phi}}$ denotes the scaling dimension of $\hat{\Phi}$ in the UV bulk theory and $C$ is a positive constant. From the second to the third line, we have used 
% In the second to last line above we used
% \al{\spl{
% \braket{l,n|\mathcal{O}_i(-1)\mathcal{O}_j(1)} 
% &= 
% \braket{l,n|\abs{\xi}^{-D}\abs{\xi}^{D}
% \mathcal{O}_i(-1)
% \abs{\xi}^{-D}\abs{\xi}^{D}
% \mathcal{O}_j(1)
% \abs{\xi}^{-D}\abs{\xi}^{D}|0}
% \\
% &=
% \braket{l,n|
% \mathcal{O}_i(-\abs{\xi})
% \mathcal{O}_j(\abs{\xi})
% |0}\abs{\xi}^{\D_i+\D_j-(\D_l+n)}\,.
% }
% \label{eq:option2}}
\begin{equation}
\abs{\xi}^{D} \mathcal{O}_i(\t) \abs{\xi}^{-D} = \abs{\xi}^{\D_i}\OO_i(\abs{\xi}\t)\, ,\qquad  \abs{\xi}^{D} \hat{\Phi}(\tau,z) \abs{\xi}^{-D} =\hat{\Phi}(\abs{\xi}\t,\abs{\xi}z)\, .
\end{equation}
% \GM{Add a comment as to how the previous to last line yields the last line.}
 The final inequality demonstrates the absolute convergence for $\abs{\xi}<1$ and implies the bound
\begin{equation}
	\begin{split}
		\abs{\braket{\hat{\Phi}(0,\xi)\mathcal{O}_i(-1)\mathcal{O}_j(1)}}\leqslant\frac{C}{(1-\abs{\xi})^{\Delta_i+\Delta_j+\Delta_{\hat{\Phi}}}}\,.
	\end{split}
\end{equation}
Similarly, the correlator instead admits a series expansion in $1/\abs{\xi}$, which converges for all $\abs{\xi}>1$. Using the unitarity condition, one can derive a similar upper bound
\begin{equation}
	\begin{split}
		\abs{\braket{\hat{\Phi}(0,\xi)\mathcal{O}_i(-1)\mathcal{O}_j(1)}}=\abs{\braket{\hat{\Phi}(0,1/\xi^*)\mathcal{O}_i(-1)\mathcal{O}_j(1)}}\leqslant\frac{C}{(1-1/\abs{\xi})^{\Delta_i+\Delta_j+\Delta_{\hat{\Phi}}}}\,.
	\end{split}
\end{equation}

What we have done above is similar to the approach in \cite{Pappadopulo:2012jk}: we establish the analyticity of the correlator by assuming OPE convergence in the Euclidean region. The analysis shows that the correlator, written as the series \eqref{eq:correlator as a series}, is uniformly bounded on any compact subset of the complex $\xi$-domains $\{|\xi|>1\}$ or $\{|\xi|<1\}$. In the Euclidean subregion of these domains, where $\xi>0$ and $\xi\neq 1$, the series can be interpreted as the decomposition of the inner product of two states in a Hilbert-space basis, and it converges absolutely. It then follows that \eqref{eq:correlator as a series} converges absolutely and uniformly for $\xi\in\mathbb{C}$ with $|\xi|\neq 1$.

Combining this uniform convergence with the fact that the correlator is analytic term-wise in the block expansion, we conclude that it is analytic in the (universal covering of the) complex domain (see e.g. \cite{ProofWikiUniformLimitAnalytic})
\begin{equation}\label{rho:domain}
	\begin{split}
		\left\{\rho\in \mathbb{C}\,|\,\text{Re}(\rho)>0,\ \rho\neq1\right\}\cup\left\{\rho\in \mathbb{C}\,|\,\text{Re}(\rho)<0,\ \rho\neq-1\right\}\,.
	\end{split}
\end{equation}
The points $\rho=\pm1$ are excluded because they correspond to $\xi=0$ and $\xi=\infty$, where the correlator develops a power-law singularity and vanishes respectively. Moreover, in the special configuration \eqref{config:special}, the full correlator and $\mathcal{G}^{\hat{\Phi}}_{ij}(\rho)$ differ only by a constant factor $2^{\Delta_i+\Delta_j}$. We thus conclude that
\begin{itemize}
	\item $\mathcal{G}^{\hat{\Phi}}_{ij}(\rho)$ is analytic in the (universal covering of) complex domain \eqref{rho:domain}; 
	\item $\mathcal{G}^{\hat{\Phi}}_{ij}(\rho)$ satisfies the upper bound
	\begin{equation}\label{Grho:bound1}
		\begin{split}
			\abs{\mathcal{G}^{\hat{\Phi}}_{ij}(\rho)}\leqslant\begin{cases}
				\frac{C'}{(1-\abs{\xi})^{\Delta_i+\Delta_j+\Delta_{\hat{\Phi}}}}, &\quad \abs{\xi}<1, \\[4pt]
				\frac{C'}{(1-1/\abs{\xi})^{\Delta_i+\Delta_j+\Delta_{\hat{\Phi}}}}, &\quad  \abs{\xi}>1,\\
			\end{cases}
		\end{split}
	\end{equation}
\end{itemize}
where the constant $C'$ is different from $C$ because of the prefactor appearing in \eqref{bkbdbd:generalform}.

\subsubsection{Analyticity from equal-time quantization}\label{subsec:equaltime}

In the equal-time quantization, one can choose an appropriate basis of bulk and boundary operators such that Hermitian conjugation acts as
\begin{equation}
	\begin{split}
		\hat{\Phi}(\tau,z)^\dagger=\hat{\Phi}(-\tau,z),\qquad\mathcal{O}_i(\tau)^\dagger=\mathcal{O}_i(-\tau)\,.
	\end{split}
\label{eq:operator conjugation in equal time}
\end{equation}

Using the unitarity condition (positivity of the norm) together with translation invariance, it follows from the argument of Osterwalder-Schrader \cite{Osterwalder:1973dx} that the spectrum of the time-translation generator is non-negative, i.e. $P\geqslant0$.\footnote{For general quantum field theory in Euclidean flat space, a polynomial bound on the correlator at large time separation is also needed for the OS argument. But here the bound follows from the OPE.} Then it follows from the argument of Glaser \cite{Glaser:1974hy} that a general bulk correlator
\begin{equation}
	\begin{split}
		\braket{\hat{\Phi}(\tau_1,z_1)\hat{\Phi}(\tau_2,z_2)\ldots\hat{\Phi}(\tau_n,z_n)}\, ,
	\end{split}
\end{equation}
admits an analytic continuation to the domain of complex time coordinates $\tau_i$ satisfying
\begin{equation}
	\begin{split}
		\text{Re}(\tau_1)>\text{Re}(\tau_2)>\ldots>\text{Re}(\tau_{n})\,.
	\end{split}
\end{equation}
This continuation coincides with the Euclidean correlator when all imaginary parts of the time variables vanish. The same conclusion holds when the bulk operators are non-identical, or when some of the bulk operators are replaced by boundary operators.

We now consider the bulk–boundary–boundary correlator in \eqref{bkbdbd:generalform} at the configuration
\begin{equation}\label{config:equaltime_special}
	\begin{split}
		(\tau,z)=(0,1),\quad \tau_1=-(\epsilon+it),\quad \tau_2=\epsilon+it\, ,\qquad (\epsilon>0)\,.
	\end{split}
\end{equation}
From the above argument, the correlator is analytic in the time variables within a complex neighborhood of this configuration. Using \eqref{def:rho}, the corresponding $\rho$ variable is
\begin{equation}
	\begin{split}
		\rho=\frac{1-(\epsilon+it)^2}{1+(\epsilon+it)^2}\,.
	\end{split}
\end{equation}
Since the image of $(\epsilon+it)^2$ is $\mathbb{C}\backslash(-\infty,0]$, the image of $\rho$ is $\mathbb{C}\backslash\big((- \infty,-1]\cup[1,+\infty)\big)$
% \begin{equation}
% 	\mathbb{C}\backslash\big((- \infty,-1]\cup[1,+\infty)\big)\,,
% \end{equation}
which precisely covers the region not reached in the radial quantization picture, namely
\begin{equation}
	\begin{split}
		\text{Re}(\rho)=0\,.
	\end{split}
\end{equation}
At the configuration \eqref{config:equaltime_special}, the full correlator and $\mathcal{G}^{\hat{\Phi}}_{ij}(\rho)$ differ by a factor $(2(\epsilon+it))^{\Delta_i+\Delta_j}$, which is analytic within the established domain. Therefore, we conclude that
\begin{itemize}
	\item In a unitary QFT on AdS$_2$, the function $\mathcal{G}^{\hat{\Phi}}_{ij}(\rho)$ is analytic in the domain
	\begin{equation}\label{rho:domain_equal_time}
		\begin{split}
			\rho\in\mathbb{C}\backslash\big((- \infty,-1]\cup[1,+\infty)\big)\,.
		\end{split}
	\end{equation}
\end{itemize}

\subsubsection{A uniform bound}\label{subsec:uniformbound}

Equation \eqref{Grho:bound1} provides a power-law bound on the correlator $\mathcal{G}^{\hat{\Phi}}_{ij}(\rho)$ when expressed as a function of $\xi\equiv\sqrt{\tfrac{1-\rho}{1+\rho}}$. Translating this bound into the $\rho$ variable at large $|\rho|$, also taking into account that the correlator is finite in any bounded region of $\rho$, we obtain
\begin{equation}
	\begin{split}
		\abs{\mathcal{G}^{\hat{\Phi}}_{ij}(\rho=Re^{i\theta})}\leqslant C_\theta (1+R)^{\Delta_i+\Delta_j+\Delta_{\hat{\Phi}}},
	\end{split}
\end{equation}
where the coefficient $C_\theta$ remains finite for any $\theta\neq\frac{\pi}{2}$.\footnote{Notice that $(1-\abs{\xi})^{-1} =\frac{R}{\cos \theta } + O(1)$ at large $R$.} In other words, the correlator exhibits the same power-law growth provided that $\rho$ does not approach infinity along the imaginary axis. Recall that $\rho$ lying on the imaginary axis corresponds to $\abs{\xi}=1$, where the bound \eqref{Grho:bound1} diverges.

From the discussion of section \ref{subsec:equaltime}, we know that $\mathcal{G}^{\hat{\Phi}}_{ij}(\rho)$ is analytic in a complex neighborhood of the imaginary axis, and it is therefore natural to expect that, even when $\rho$ tends to infinity along the imaginary direction, the correlator remains bounded by the same power-law behaviour. In other words, one might conjecture the existence of a uniform power-law bound:
\begin{equation}\label{Grho:bound2}
	\begin{split}
		\abs{\mathcal{G}^{\hat{\Phi}}_{ij}(\rho=Re^{i\theta})}\stackrel{?}{\leqslant} C (1+R)^{\Delta_i+\Delta_j+\Delta_{\hat{\Phi}}},\qquad C<+\infty\,.
	\end{split}
\end{equation}
% \GM{The spacing here is all fucked because I forced the footnote 6 to not be spread over 2 pages. Not sure how to solve this now. Are we fine with a footnote over 2 pages? }
However, this property does not hold in general for arbitrary analytic functions.\footnote{
For a counterexample, consider the following entire function \cite{newman1976entire}:
\begin{equation*}
    H(z)=\int_0^{\infty} dt\, t^{-t} e^{z t}\,.
\end{equation*}
This function is bounded along any straight line through the origin except for the real axis, but it grows very rapidly along the positive real axis. The counterexample is then given by $g(\rho)=\left(1-\rho^2\right)^{\tfrac{1}{2}(\Delta_i+\Delta_j+\Delta_{\hat\Phi})}H(i\rho)$.
}
For this reason, we introduce a mild technical assumption on the growth of the correlator.

\noindent\textbf{Assumption:} In the domain \eqref{rho:domain_equal_time}, the correlator satisfies 
\begin{equation}\label{Grho:assumption}
	\begin{split}
		\abs{\mathcal{G}^{\hat{\Phi}}_{ij}(\rho)}\leqslant A e^{B\abs{\rho}^M} 
	\end{split}
\end{equation}
for some finite constants $A$, $B$, and $M$. 

This assumption is very mild, since the limit $\rho\rightarrow\infty$ corresponds to $\xi\rightarrow \pm i$, which is the light-cone singularity of the correlator. The light-cone singularity of a QFT correlator is generally expected to exhibit power-law rather than exponential behaviour. Nevertheless, it remains an interesting question whether this assumption can be removed, which we leave for future study.

With the extra assumption, \eqref{Grho:bound2} now follows as a direct consequence of the following theorem.
\begin{theorem}\label{theorem:PL}
	(\textbf{Phragmén–Lindelöf theorem, sector version})  
	Let $f(z)$ be an analytic function in the sector domain
	\begin{equation}\label{def:sector}
		\mathcal{S}_{\theta_1,\theta_2}:=\left\{\,z=re^{i\theta}\ \big|\ r>0,\ \theta_1<\theta<\theta_2\,\right\}\,,
	\end{equation}
    and continuous up to its boundary. Suppose that $f(z)$ is uniformly bounded along the boundary of the sector:
	\begin{equation}\label{f:boundM}
		\abs{f\!\left(re^{i\theta_1}\right)},\ \abs{f\!\left(re^{i\theta_2}\right)}\leqslant C,\qquad \forall\,r\geqslant0\,,
	\end{equation}
	and that, as $r\rightarrow\infty$, 
	\begin{equation}
		\abs{f(z)}\leqslant A e^{\abs{z}^\beta}\,,
	\end{equation}
	uniformly in $\arg(z)$ for some finite $A$ and some $\beta<\tfrac{\pi}{\theta_2-\theta_1}$.  
	Then
	\begin{equation}
		\begin{split}
			\abs{f(z)}\leqslant C\, ,
		\end{split}
	\end{equation}
	for all $z\in\mathcal{S}_{\theta_1,\theta_2}$.
\end{theorem}
We leave the proof of theorem~\ref{theorem:PL} to appendix~\ref{app:PL}. For the discussion of more general versions of Phragmén–Lindelöf theorem, see~\cite[section~5.6]{titchmarsh1939theory}.

Let us now show how the uniform bound \eqref{Grho:bound2} follows near $\theta=\tfrac{\pi}{2}$; the argument for $\theta=-\tfrac{\pi}{2}$ is analogous.  
Consider the sector $\mathcal{S}_{\theta_1,\theta_2}$ with $\theta_1<\tfrac{\pi}{2}<\theta_2$, and choose $\theta_1$ and $\theta_2$ sufficiently close to $\tfrac{\pi}{2}$ such that
\begin{equation}
	\begin{split}
		\theta_2-\theta_1<\frac{\pi}{M}\,.
	\end{split}
\end{equation}
One can verify that the conditions of theorem~\ref{theorem:PL} are satisfied for
\begin{equation}
	\begin{split}
		f(z)=(i+z)^{-(\Delta_i+\Delta_j+\Delta_{\hat{\Phi}})}\,
		\mathcal{G}^{\hat{\Phi}}_{ij}(z)\,.
	\end{split}
\end{equation}
Applying the theorem to this function $f(z)$ yields the desired uniform bound \eqref{Grho:bound2} near $\theta=\tfrac{\pi}{2}$.

\section{Locality constraints}
\label{sec:locality constraints}

%\GM{In this section, we derive a new conformal block expansion assuming analyticity and a polynomial bound. For unitary theories, we proved this in the previous section. Adapt this sentence}\ML{Attempt:}

In the previous section, using unitarity in two different quantization pictures, we established the domain of analyticity and a power-law bound for bulk-boundary-boundary correlators in AdS$_2$.  In this section, we use these facts to prove the existence and convergence properties of local block decompositions for these correlators, as well as sum rules that the QFT data must satisfy. 
While for unitary theories, we could prove these requirements rigorously, these results hold also in non-unitary theories, provided that correlators enjoy the same analiticity domain, power-law bound, and convergence property of the conformal block expansion.

\subsection{Dispersion relation for the full correlator}

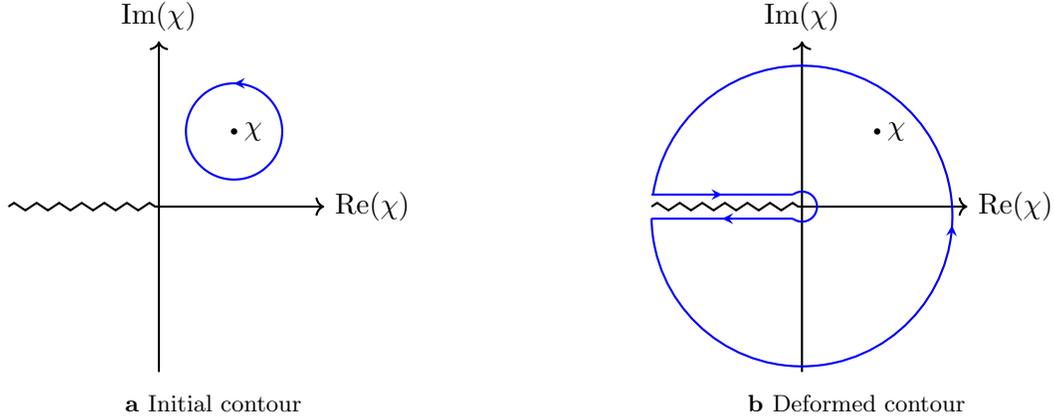
\begin{figure}[t]
	\centering
	\begin{subfigure}[b]{0.45\textwidth}
		\centering
		\begin{tikzpicture}[scale=0.4,baseline={(0,0)}]
			% Draw axes
			\draw[->,thick] (0,0) -- (5.5,0) node[right] {Re$(\chi)$};
			\draw[->,thick] (0,-5.5) -- (0,5.5) node[above] {Im$(\chi)$};
			
			% Mark the center point
			\fill (2.5,2.5) circle (3pt) node[right] {$\chi$};
			
			% Draw the initial circular contour with arrow
			\draw[blue,thick,decoration={markings, mark=at position 0.25 with {\arrow{stealth}}},postaction={decorate}] 
			(2.5,2.5) circle (1.6);
			
			% Branch cut - zigzag line from -5 to 0
			\draw[thick, decorate, decoration={zigzag, amplitude=0.5mm, segment length=3mm}] 
			(-5,0) -- (0,0);
		\end{tikzpicture}
		\caption{Initial contour}
		\label{fig:initial}
	\end{subfigure}
	\hfill
	\begin{subfigure}[b]{0.45\textwidth}
		\centering
		\begin{tikzpicture}[scale=0.4,baseline={(0,0)}]
			% Draw axes
			\draw[->,thick] (0,0) -- (5.5,0) node[right] {Re$(\chi)$};
			\draw[->,thick] (0,-5.5) -- (0,5.5) node[above] {Im$(\chi)$};
			
			% Mark the center point
			\fill (2.5,2.5) circle (3pt) node[right] {$\chi$};
			
			% Draw the deformed contour
			% Straight line from (-5,0.4) to (-0.3,0.4)
			\draw[blue,thick,decoration={markings, mark=at position 0.5 with {\arrow{stealth}}},postaction={decorate}] 
			(-5,0.4) -- (-0.3,0.4);
			
			% Small arc centered around zero, radius 0.5, from (-0.3,0.4) to (-0.3,-0.4)
			\draw[blue,thick] 
			(-0.3,0.4) arc[start angle=126.87, end angle=-126.87, radius=0.5];
			
			% Straight line from (-0.3,-0.4) to (-5,-0.4)
			\draw[blue,thick,decoration={markings, mark=at position 0.5 with {\arrow{stealth}}},postaction={decorate}] 
			(-0.3,-0.4) -- (-5,-0.4);
			
			% Big arc centered around zero, connecting (-5,-0.4) and (-5,0.4)
			\draw[blue,thick,decoration={markings, mark=at position 0.5 with {\arrow{stealth}}},postaction={decorate}] 
			(-5,-0.4) arc[start angle=-179, end angle=172, radius=5];
			
			% Branch cut - zigzag line from -5 to 0
			\draw[thick, decorate, decoration={zigzag, amplitude=0.5mm, segment length=3mm}] 
			(-5,0) -- (0,0);
		\end{tikzpicture}
		\caption{Deformed contour}
		\label{fig:deformed}
	\end{subfigure}
	\caption{Contour deformation.}
	\label{fig:contour-deformation}
\end{figure}

In the previous section, we established the analyticity domain of $\mathcal{G}^{\hat{\Phi}}_{ij}(\rho)$ using two different quantization pictures. The domain \eqref{rho:domain_equal_time} obtained from the equal-time quantization includes a complex neighborhood of $\rho=0$. This allows us to decompose $\mathcal{G}^{\hat{\Phi}}_{ij}(\rho)$ appearing in \eqref{bkbdbd:generalform} into its even and odd parts:
\begin{equation}\label{Gbkbdbd:evenodd}
	\begin{split}
		\mathcal{G}^{\hat{\Phi}}_{ij}(\rho)
		= [\mathcal{G}^{\hat{\Phi}}_{ij}]_e(\chi)
		+ \rho\, [\mathcal{G}^{\hat{\Phi}}_{ij}]_o(\chi)\,,
	\end{split}
\end{equation}
where $\chi=1-\rho^2$ (see \eqref{eq:chifromrho}) and 
\begin{equation}\label{def:G01}
	\begin{split}
		[\mathcal{G}^{\hat{\Phi}}_{ij}]_e(\chi)&:=\frac{1}{2}\left[\mathcal{G}^{\hat{\Phi}}_{ij}(\rho)+\mathcal{G}^{\hat{\Phi}}_{ij}(-\rho)\right]\,, \\
		[\mathcal{G}^{\hat{\Phi}}_{ij}]_o(\chi)&:=\frac{1}{2\rho}\left[\mathcal{G}^{\hat{\Phi}}_{ij}(\rho)-\mathcal{G}^{\hat{\Phi}}_{ij}(-\rho)\right]\,.
	\end{split}
\end{equation}
According to \eqref{rho:domain_equal_time}, both $[\mathcal{G}^{\hat{\Phi}}_{ij}]_e(\chi)$ and $[\mathcal{G}^{\hat{\Phi}}_{ij}]_o(\chi)$ are analytic functions of $\chi$ in the domain
\begin{equation}\label{chi:domain_equal_time}
	\begin{split}
		\chi\in\mathbb{C}\backslash(-\infty,0]\, .
	\end{split}
\end{equation}
We may therefore divide $[\mathcal{G}^{\hat{\Phi}}_{ij}]_{e/o}(\chi)$ by a factor $\chi^{\alpha}$ and apply Cauchy’s theorem:
\begin{equation}\label{G:cauchy}
	\begin{split}
		[\mathcal{G}^{\hat{\Phi}}_{ij}]_{e/o}(\chi)
		&=\chi^\alpha\oint\frac{d\chi'}{2\pi i}
		\frac{\chi'^{-\alpha}}{\chi'-\chi}
		[\mathcal{G}^{\hat{\Phi}}_{ij}]_{e/o}(\chi')\,.
	\end{split}
\end{equation}
The closed integration contour lies within the domain \eqref{chi:domain_equal_time}, see figure \ref{fig:initial}.

We now show that, for sufficiently large real positive $\alpha$, the contour in \eqref{G:cauchy} can be deformed into the contour shown in figure \ref{fig:deformed}. If dropping the contour at infinity is allowed, this deformation leads to an integral along the negative real axis
\begin{equation}\label{G:dispersion}
	\begin{split}
		[\mathcal{G}^{\hat{\Phi}}_{ij}]_{e/o}(\chi)
		&=\chi^\alpha\int_{-\infty}^{0}\frac{d\chi'}{2\pi i}
		\frac{1}{\chi'-\chi}
		\text{Disc}\!\left(\chi'^{-\alpha} [\mathcal{G}^{\hat{\Phi}}_{ij}]_{e/o}(\chi')\right),
		\qquad \chi\notin(-\infty,0]\,,
	\end{split}
\end{equation}
where the discontinuity is defined as
\begin{equation}\label{def:DiscG}
	\begin{split}
		\text{Disc}\!\left([\mathcal{G}^{\hat{\Phi}}_{ij}]_{e/o}(\chi)\right)
		:=\lim_{\epsilon\rightarrow0^+}
		\Big([\mathcal{G}^{\hat{\Phi}}_{ij}]_{e/o}(\chi+i\epsilon)
		-[\mathcal{G}^{\hat{\Phi}}_{ij}]_{e/o}(\chi-i\epsilon)\Big)\,.
	\end{split}
\end{equation}
After the contour deformation, an additional small contour must be included around $\chi=0$.  
This subtraction removes any potential power-law divergence near $\chi=0$, ensuring that the integral in \eqref{G:dispersion} is well defined for sufficiently large $\alpha$.

The deformation from \eqref{G:cauchy} to \eqref{G:dispersion} requires that the contour at infinity gives no contribution, i.e.,
\begin{equation}\label{eq:InfContour}
	\begin{split}
		\lim\limits_{R\rightarrow\infty}\int_{-\pi}^{\pi}\frac{d\theta}{2\pi}
		\frac{R^{1-\alpha}}{Re^{i\theta}-\chi}
		[\mathcal{G}^{\hat{\Phi}}_{ij}]_{e/o}(Re^{i\theta})=0\,.
	\end{split}
\end{equation}

In section \ref{subsec:uniformbound}, we established the uniform bound \eqref{Grho:bound2} for the full correlator $\mathcal{G}^{\hat{\Phi}}_{ij}(\rho)$, assuming the growth condition \eqref{Grho:assumption}. Using \eqref{def:chi} and \eqref{def:G01}, we then obtain the corresponding bounds for its even and odd parts:
\begin{equation}
	\begin{split}
		[\mathcal{G}^{\hat{\Phi}}_{ij}]_e(\chi)&\leqslant C_e\,\abs{\chi}^{\tfrac{1}{2}(\Delta_i+\Delta_j+\Delta_{\hat{\Phi}})}\,, \\
		[\mathcal{G}^{\hat{\Phi}}_{ij}]_o(\chi)&\leqslant C_o\,\abs{\chi}^{\tfrac{1}{2}(\Delta_i+\Delta_j+\Delta_{\hat{\Phi}}-1)}\,. \\
	\end{split}
\end{equation}
Therefore, discarding the contour at infinity requires
\begin{equation}\label{alpha:range}
	\begin{split}
		\alpha>\begin{cases}
			\frac{1}{2}(\Delta_i+\Delta_j+\Delta_{\hat{\Phi}}), & \text{for }[\mathcal{G}^{\hat{\Phi}}_{ij}]_e, \\[4pt]
			\frac{1}{2}(\Delta_i+\Delta_j+\Delta_{\hat{\Phi}}-1), & \text{for }[\mathcal{G}^{\hat{\Phi}}_{ij}]_o. \\
		\end{cases}
	\end{split}
\end{equation}
We thus arrive at the following conclusion:
\begin{center}
\noindent\fbox{%
  \begin{minipage}{0.95\textwidth} % Adjust width as needed
    In a unitary QFT on AdS$_2$ with a conformal boundary condition, the dispersion relation \eqref{G:dispersion} holds for any bulk–boundary–boundary correlator satisfying the bound \eqref{Grho:assumption}, provided that $\alpha$ lies within the range \eqref{alpha:range}.
  \end{minipage}%
}\end{center}

\subsection{Local block expansion}

In the right-hand side of \eqref{G:dispersion}, we can expand the discontinuity in terms of conformal blocks. The resulting series is uniformly convergent for $\chi\in(-\infty,0]$, and for $\alpha$ within the range \eqref{alpha:range} the integral absolutely converges, thus by the dominated convergence theorem we can interchange the order of summation and integration. 
% \cite{Qiao:2017lkv,Levine:2023ywq}.\JQ{I think there's no need to cite these two papers here.}
This leads to the following new expansion:
\begin{equation}\label{Gbkbdbd:newexpansion}
	\begin{split}
		[\mathcal{G}^{\hat{\Phi}}_{ij}]_{e}(\chi)
		&=\sum_{l}b^{\hat{\Phi}}_{l}\,
		\frac{C_{ijl}+C_{jil}}{2}\,
		G_{\Delta_l}^{\Delta_i\Delta_j,(\alpha)}(\chi)\,, \\
		[\mathcal{G}^{\hat{\Phi}}_{ij}]_{o}(\chi)
		&=\sum_{l}b^{\hat{\Phi}}_{l}\,
		\frac{C_{jil}-C_{ijl}}{2}\,
		\tilde{G}_{\Delta_l}^{\Delta_i\Delta_j,(\beta)}(\chi)\,,
	\end{split}
\end{equation}
where the functions $G_{\Delta_l}^{\Delta_i\Delta_j,(\alpha)}(\chi)$ and $\tilde{G}_{\Delta_l}^{\Delta_i\Delta_j,(\beta)}(\chi)$ are the \emph{local blocks} corresponding to the even and odd parts of the full correlator, defined by
\begin{equation}\label{def:Gbkbdbd_local}
	\begin{split}
		G^{\Delta_i\Delta_j,(\alpha)}_{\Delta_l}(\chi)
		&:=\chi^\alpha\sin\!\left[\tfrac{\pi}{2}(\Delta_l-2\alpha)\right]
		\!\int_{-\infty}^{0}\!\frac{d\chi'}{\pi}
		\frac{(-\chi')^{\Delta_l/2-\alpha}}{\chi'-\chi}
		\hyperF{\tfrac{\Delta_{l}+\D_{ij}}{2}}{\tfrac{\Delta_{l}+\D_{ji}}{2}}{\Delta_l+\tfrac{1}{2}}{\chi'}\,, \\
		\tilde{G}^{\Delta_i\Delta_j,(\alpha)}_{\Delta_l}(\chi)
		&:=\chi^\alpha\sin\!\left[\tfrac{\pi}{2}(\Delta_l-2\alpha)\right]
		\!\int_{-\infty}^{0}\!\frac{d\chi'}{\pi}
		\frac{(-\chi')^{\Delta_l/2-\alpha}}{(\chi'-\chi)\sqrt{1-\chi'}}
		\hyperF{\tfrac{\Delta_{l}+\D_{ij}}{2}}{\tfrac{\Delta_{l}+\D_{ji}}{2}}{\Delta_l+\tfrac{1}{2}}{\chi'}\,.
	\end{split}
\end{equation}
The explicit form of $G^{\Delta_i\Delta_j,(\alpha)}_{\Delta_l}(\chi)$ is known~\cite{Levine:2023ywq,Meineri:2023mps}:
\begin{equation}\label{localblock:bkbdbd_even}
	\begin{split}
		G^{\Delta_i\Delta_j,(\alpha)}_{\Delta_l}(\chi)
		&=
        G^{\Delta_i\Delta_j}_{\Delta_l}(\chi)
        -\frac{\Gamma\!\left(\Delta_l+\frac{1}{2}\right)
			\Gamma\!\left(\frac{\Delta_{ij}}{2}+\alpha\right)
			\Gamma\!\left(\frac{\Delta_{ji}}{2}+\alpha\right)}
		{\Gamma\!\left(\frac{\Delta_{l}+\Delta_{ij}}{2}\right)
			\Gamma\!\left(\frac{\Delta_{l}+\Delta_{ji}}{2}\right)
			\Gamma\!\left(\frac{\Delta_l+1}{2}+\alpha\right)
			\Gamma\!\left(-\frac{\Delta_l}{2}+\alpha+1\right)} \\
		&\hspace{5cm}\times \chi^\alpha
		\pFq{3}{2}{1,\frac{\Delta_{ij}}{2}+\alpha,\frac{\Delta_{ji}}{2}+\alpha}
		{\frac{\Delta_l+1}{2}+\alpha,-\frac{\Delta_l}{2}+\alpha+1}{\chi}\,,
	\end{split}
\end{equation}
where the normal block $ G^{\Delta_i\Delta_j}_{\Delta_l}(\chi)$ is given in \eqref{def:bkbdbd block}. The function $\tilde{G}^{\Delta_i\Delta_j,(\alpha)}_{\Delta_l}(\chi)$ does not appear to have been discussed in the existing literature. However, using in the second line of \eqref{def:Gbkbdbd_local} the hypergeometric identity
\begin{equation}
	\frac{1}{\sqrt{1-\chi}}
	\hyperF{\tfrac{\Delta_{l}+\D_{ij}}{2}}{\tfrac{\Delta_{l}+\D_{ji}}{2}}{\Delta_l+\tfrac{1}{2}}{\chi}
	=\hyperF{\tfrac{\Delta_{l}+\D_{ij}+1}{2}}{\tfrac{\Delta_{l}+\D_{ji}+1}{2}}{\Delta_l+\tfrac{1}{2}}{\chi}\,,
\end{equation} 
the same method used to compute $G^{\Delta_i\Delta_j,(\alpha)}_{\Delta_l}(\chi)$ can also be applied to $\tilde{G}^{\Delta_i\Delta_j,(\alpha)}_{\Delta_l}(\chi)$. The final result is\footnote{Note that if we choose to write ${G}^{\Delta_i\Delta_j,(\alpha)}_{\Delta_l}(\chi)$ such that all the $\D_{ij}$ and $\D_{ji}$ have $+$ sign in front, as in \eqref{localblock:bkbdbd_even}, then $\tilde{G}^{\Delta_i\Delta_j,(\alpha)}_{\Delta_l}(\chi)$ can be simply obtained by sending $(\D_{ij},\D_{ji})\to(\D_{ij}+1,\D_{ji}+1)$ in ${G}^{\Delta_i\Delta_j,(\alpha)}_{\Delta_l}(\chi)$.}
\begin{equation}\label{localblock:bkbdbd_odd}
	\begin{split}
		\tilde{G}^{\Delta_i\Delta_j,(\alpha)}_{\Delta_l}(\chi)
		&=\frac{G^{\Delta_i\Delta_j}_{\Delta_l}(\chi)}{\sqrt{1-\chi}}
  %       \,
		% {}_2F_1\!\left(\tfrac{\Delta_{lij}}{2},\tfrac{\Delta_{lji}}{2};
		% \Delta_l+\tfrac{1}{2};\chi\right) 
 -\frac{\Gamma(\Delta_l+\tfrac{1}{2})
			\Gamma(\tfrac{\Delta_{ij}+1}{2}+\alpha)
			\Gamma(\tfrac{\Delta_{ji}+1}{2}+\alpha)}
		{\Gamma(\tfrac{\Delta_{l}+\Delta_{ij}+1}{2})
			\Gamma(\tfrac{\Delta_{l}+\Delta_{ji}+1}{2})
			\Gamma(\tfrac{\Delta_l+1}{2}+\alpha)
			\Gamma(1-\tfrac{\Delta_l}{2}+\alpha)} \\
		&\hspace{5cm}\times \chi^\alpha
		\pFq{3}{2}{1,\tfrac{\Delta_{ij}+1}{2}+\alpha,\tfrac{\Delta_{ji}+1}{2}+\alpha}
		{\tfrac{\Delta_l+1}{2}+\alpha,-\tfrac{\Delta_l}{2}+\alpha+1}{\chi}\,.
	\end{split}
\end{equation}
We refer to $G^{\Delta_i\Delta_j,(\alpha)}_{\Delta_l}(\chi)$ and $\tilde{G}^{\Delta_i\Delta_j,(\alpha)}_{\Delta_l}(\chi)$ as the \emph{even} and \emph{odd local blocks}, respectively, since they contribute to the even and odd parts of the full correlator $\mathcal{G}^{\hat{\Phi}}_{ij}(\rho)$. The final result of the local block expansion is
\al{
\GG(\r)=\sum_l b^{\hat\F}_l \frac{C_{ijl}+C_{jil}}{2}G^{\Delta_i\Delta_j,(\a)}_{\D_l}(\chi)+
\r \sum_l b^{\hat\F}_l \frac{C_{ijl}-C_{jil}}{2}\tilde{G}^{\Delta_i\D_j,(\b)}_{\D_l}(\chi)\,,
}
where the odd part is multiplied by  $\rho$.

\subsection{Locality sum rules}

The local blocks \eqref{localblock:bkbdbd_even} and \eqref{localblock:bkbdbd_odd} can be rewritten as sums of standard conformal blocks:
\begin{equation}\label{localblock:rewrite}
	\begin{split}
		G^{\Delta_i\Delta_j,(\alpha)}_{\Delta_l}(\chi)
		&=G^{\Delta_i\Delta_j}_{\Delta_l}(\chi)
		-\sum_{n=0}^{\infty}\theta^{(\alpha)}_n(\Delta_l;\Delta_{ij},\Delta_{ji})G^{\Delta_i\Delta_j}_{2\alpha+2n}(\chi)\,, \\
		\tilde{G}^{\Delta_{ij},\Delta_{ji},(\alpha)}_{\Delta_l}(\chi)
		&=\frac{1}{\sqrt{1-\chi}}\left[G^{\Delta_i\Delta_j}_{\Delta_l}(\chi)
		-\sum_{n=0}^{\infty}\theta^{(\alpha)}_n(\Delta_l;\Delta_{ij}+1,\Delta_{ji}+1)G^{\Delta_i\Delta_j}_{2\alpha+2n}(\chi)\right]\,, \\
	\end{split}
\end{equation}
where the coefficient function $\theta^{(\alpha)}_n$ is defined by\footnote{The function $\q_n^{(\a)}$ is slightly different from that in \cite[eq.(3.27)]{Levine:2023ywq}. We have set dimension $d=1$ and $\tilde\alpha_{\text{theirs}}=\alpha_{\text{ours}}-1$, $n_{\rm theirs}=n_{\rm ours}+1$.}
\begin{equation}\label{def:theta}
	\begin{split}
		\theta^{(\alpha)}_n(\Delta_l;a,b)&:=\frac{4 (-1)^n }{n!\,(2 \alpha +\Delta_l+2 n-1) (2
			\alpha +2n-\Delta_l)} \\
		&\quad\times\frac{\Gamma \left(\Delta_l+\frac{1}{2}\right) \Gamma \left(n+2 \alpha
			-\frac{1}{2}\right) \Gamma \left(\frac{a}{2}+n+\alpha \right) \Gamma
			\left(\frac{b}{2}+n+\alpha \right)}{\Gamma \left(\frac{a+\Delta_l}{2}\right)
			\Gamma \left(\frac{b+\Delta_l}{2}\right) \Gamma \left(\frac{\Delta_l-1}{2}+\alpha
			\right) \Gamma \left(\alpha -\frac{\Delta_l}{2}\right)  \Gamma \left(2 n+2 \alpha -\frac{1}{2}\right)}\,.
	\end{split}
\end{equation}

\begin{figure}[!t]
    \centering
    \includegraphics[width=0.85\linewidth]{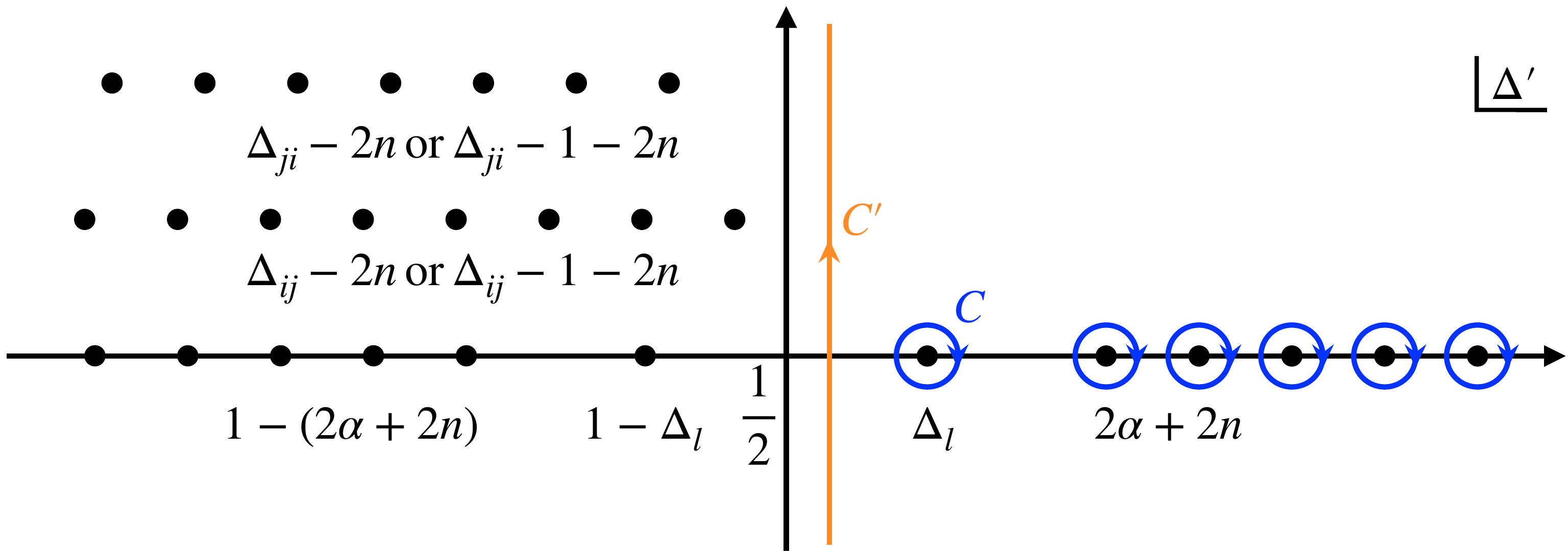}
    \caption{Integration contours $C$ (blue) and $C'$ (orange) of the integral transform \eqref{localblock:int_rep}. The poles starting from $\D_{ij}$ and $\D_{ji}$ belong to the even local block integral representation while those starting from $\D_{ij}-1$ and $\D_{ji}-1$ belong to the odd one. These poles are shifted along the imaginary axis to avoid cluttering. Notice the origin is at $\D'=\tfrac{1}{2}$. }
    \label{fig:integral transform}
\end{figure}

The local blocks also admit an integral representation in terms of standard conformal blocks:
\begin{equation}\label{localblock:int_rep}
	\begin{split}
		G^{\Delta_i\Delta_j,(\alpha)}_{\Delta_l}(\chi)&=
		\int_{C}\frac{d\Delta'}{2\pi i}\,
		K^{(\alpha)}(\Delta',\Delta_l;\Delta_{ij},\Delta_{ji})\,
		G^{\Delta_i\Delta_j}_{\Delta'}(\chi)\,, \\
		\tilde{G}^{\Delta_i\Delta_j,(\alpha)}_{\Delta_l}(\chi)
		&=\frac{1}{\sqrt{1-\chi}}
		\int_{C}\frac{d\Delta'}{2\pi i}\,
		K^{(\alpha)}(\Delta',\Delta_l;\Delta_{ij}+1,\Delta_{ji}+1)\,
		G^{\Delta_i\Delta_j}_{\Delta'}(\chi)\,, \\
	\end{split}
\end{equation}
where the kernel $K$ is defined by
\begin{equation}
	\begin{split}
		K^{(\alpha)}(\Delta',\Delta_l;a,b)
		:=\frac{1-2\Delta_l}{(\Delta_l-\Delta')(1-\Delta_l-\Delta')}
		\frac{\Gamma(\alpha-\tfrac{\Delta'}{2})
			\Gamma(\alpha-\tfrac{1-\Delta'}{2})
			\Gamma(\Delta_l-\tfrac{1}{2})
			\Gamma(\tfrac{\Delta'+a}{2})
			\Gamma(\tfrac{\Delta'+b}{2})}
		{\Gamma(\alpha-\tfrac{\Delta_l}{2})
			\Gamma(\alpha-\tfrac{1-\Delta_l}{2})
			\Gamma(\Delta'-\tfrac{1}{2})
			\Gamma(\tfrac{\Delta_l+a}{2})
			\Gamma(\tfrac{\Delta_l+b}{2})}\,.
	\end{split}
\end{equation}
The integration contour $C$ in \eqref{localblock:int_rep} is chosen such that the pole at $\Delta'=\Delta_l$ and the poles at $\Delta'=2\alpha+2n$ ($n\in\mathbb{N}_0$) are enclosed clockwise; no other poles are included (see figure \ref{fig:integral transform}). The residues of the kernel at $\Delta'=2\alpha+2n$ are the coefficients $\theta^{(\alpha)}_n$ in \eqref{localblock:rewrite}:
\begin{equation}
	\underset{\Delta'=2\alpha+2n}{\text{Res}}
	K^{(\alpha)}(\Delta',\Delta_l;a,b)
    =
    \theta^{(\alpha)}_n(\Delta_l;a,b)\,.
\end{equation}
In addition, we built the kernel such that 
\begin{equation}
\underset{\Delta'=\Delta_l}{\text{Res}}
	K^{(\alpha)}(\Delta',\Delta_l;a,b)=-1\, .
\end{equation}

It is possible and convenient for numerics (see \cite{Loparco:2026fki}) to deform the contour $C$ to a straight contour $C'$ parallel to the imaginary axis in $\D'$ plane, as shown in figure \ref{fig:integral transform}. The arc at infinity can be dropped because in the limit $|\D'|\gg1$ with $\abs{\arg{\D'}}<\pi$, the integrand behaves as\footnote{To determine the large $\D'$ behaviour of the conformal block \eqref{def:bkbdbd block}, we use the following integral representation of the hypergeometric function
\begin{equation*}
    \begin{split}
        \hyperF{a_1}{a_2}{b_1}{z}&=\frac{\Gamma (b_1)}{\Gamma (a_1) \Gamma (a_2) \Gamma (b_1-a_1) \Gamma (b_1-a_2)}
\\
&\quad\times\int \frac{ds}{2\pi i}\Gamma (s) \Gamma (a_1-s) \Gamma (a_2-s) \Gamma (-a_1-a_2+b_1+s)(1-z)^{-s}\, , \qquad (|{\rm arg}(1-z)|<\pi)\,,
    \end{split}
\end{equation*}
and do a saddle point approximation. Also, note that this hypergeometric ${}_2F_1$ is regulated by the factor $1/\G(\D'-1/2)$ in $K^{(\a)}$.}
\al{
K^{(\alpha)}(\Delta',\Delta_l;\Delta_{ij},\Delta_{ji})\,
		G^{\Delta_i\Delta_j}_{\Delta'}(\chi)
\propto
(\D')^{2 \alpha -\frac{7}{2}} \left(\frac{\sqrt{\chi }}{\sqrt{1-\chi }+1}\right)^{\D'} \csc \left(\pi  \left(\alpha -\frac{\D'}{2}\right)\right)\,,
}
and it is exponentially suppressed in the right half of the complex-$\D'$ plane. The integrand for $\tilde{G}_{\D_l}^{\D_i\D_j,(\a)}(\chi)$ has the same $\D'$ dependence, so its arc at infinity is also exponentially suppressed. In order to avoid picking up any extra poles during the contour deformation, we need to assume
\al{
\D_l > \max\left(\frac{1}{2},\D_{ij},\D_{ji}\right)
\quad \text{or}\quad 
\D_l > \max\left(\frac{1}{2},\D_{ij}-1,\D_{ji}-1\right)\, .
}
The cases for more general $\D_i,\D_j,\D_l$ can be obtained by analytic continuation. When extra poles cross the $C'$ contour, we also need to pick up their residues.

We now have two distinct bases for expanding the even and odd parts of the full correlator. For example, for the even part we can write
\begin{equation}\label{exp:difference}
	\begin{split}
		[\mathcal{G}^{\hat{\Phi}}_{ij}]_{e}(\chi)
		&=\sum_{l}b^{\hat{\Phi}}_{l}\,
		\frac{C_{ijl}+C_{jil}}{2}\,
		G_{\Delta_l}^{\Delta_i\Delta_j}(\chi)
		=\sum_{l}b^{\hat{\Phi}}_{l}\,
		\frac{C_{ijl}+C_{jil}}{2}\,
		G_{\Delta_l}^{\Delta_i\Delta_j,(\alpha)}(\chi)\,. \\
	\end{split}
\end{equation}
Consistency between these two expansions requires that their difference vanishes. Using \eqref{localblock:rewrite}, this implies
\begin{equation}
	\begin{split}
		\sum_{l}b^{\hat{\Phi}}_{l}\,
		\frac{C_{ijl}+C_{jil}}{2}\sum_{n=0}^{\infty}\theta^{(\alpha)}_n(\Delta_l;\Delta_{ij},\Delta_{ji})G^{\Delta_i\Delta_j}_{2\alpha+2n}(\chi)=0\,.
	\end{split}
\end{equation}
This sum is absolutely convergent near $\chi=0$. Expanding the left-hand side as a power series in $\chi^{\alpha+n}$, the vanishing of the coefficients at each order yields the following sum rules:
% \begin{equation}\label{sumrule:even}
% 	\begin{split}
% 		&\sum_{l}b^{\hat{\Phi}}_{l}\,
% 		\frac{C_{ijl}+C_{jil}}{2}\theta^{(\alpha)}_n(\Delta_l;\Delta_{ij},\Delta_{ji})=0, \\
% 		&\text{for all }n\in\mathbb{N}_0\ \text{and }\alpha>\frac{1}{2}(\Delta_i+\Delta_j+\Delta_{\hat{\Phi}})\,.
% 	\end{split}
% \end{equation}

\begin{equation}\label{sumrule:even}
	\boxed{	\sum_{l}b^{\hat{\Phi}}_{l}\,
		\frac{C_{ijl}+C_{jil}}{2}\theta^{(\alpha)}_n(\Delta_l;\Delta_{ij},\Delta_{ji})=0, 
        \quad\forall \,n\in\mathbb{N}_0\ \text{and }\alpha>\frac{1}{2}(\Delta_i+\Delta_j+\Delta_{\hat{\Phi}})\,.}
\end{equation}
Applying the same reasoning to the odd part of the correlator gives
% \begin{equation}\label{sumrule:odd}
% 	\begin{split}
% 		&\sum_{l}b^{\hat{\Phi}}_{l}\,
% 		\frac{C_{ijl}-C_{jil}}{2}\theta^{(\alpha)}_n(\Delta_l;\Delta_{ij}+1,\Delta_{ji}+1)=0, \\
% 		&\text{for all }n\in\mathbb{N}_0\ \text{and }\alpha>\frac{1}{2}(\Delta_i+\Delta_j+\Delta_{\hat{\Phi}}-1)\,.
% 	\end{split}
% \end{equation}
\begin{equation}\label{sumrule:odd}
		\boxed{\sum_{l}b^{\hat{\Phi}}_{l}\,
		\frac{C_{ijl}-C_{jil}}{2}\theta^{(\alpha)}_n(\Delta_l;\Delta_{ij}+1,\Delta_{ji}+1)=0, \quad \forall \,n\in\mathbb{N}_0\ \text{and }\alpha>\frac{1}{2}(\Delta_i+\Delta_j+\Delta_{\hat{\Phi}}-1)\,.}
\end{equation}
These sum rules constitute the main result of this paper.

\medskip

Let us comment on the essential difference between AdS$_2$ and higher-dimensional AdS. In higher dimensions, i.e.\ AdS$_{d+1}$ with $d\geqslant2$, there is no variable $\rho$, and $\chi$ serves as the fundamental $SO^+(1,d)$ invariant. The function $\mathcal{G}^{\hat{\Phi}}_{ij}$ depends directly on $\chi$ and admits a Taylor expansion in $\chi-1$ near $\chi=1$.  
Consistently, for the boundary CFT$_d$, the scalar OPE coefficients $C_{ijk}$ are invariant under all permutations of the indices.  

In contrast, in AdS$_2$, the fundamental $SO^+(1,2)$ invariant is $\rho$ rather than $\chi$.  
The function $\mathcal{G}^{\hat{\Phi}}_{ij}(\rho)$ can differ for $\rho$ and $-\rho$ (which correspond to the same $\chi=1-\rho^2$), implying that the natural expansion parameter is $\sqrt{1-\chi}$.  
Moreover, for CFT$_1$, the OPE coefficients $C_{ijk}$ are invariant only under cyclic (even) permutations but not under odd permutations.  
This distinction explains why there exists only one type of locality sum rule in higher dimensions, whereas two independent types arise in AdS$_2$.

\subsection{Comments on conformal line defects}
This work primarily focuses on QFTs in AdS$_2$. However, the arguments presented here also apply to CFT$_d$ with a conformal line defect ($d>2$) or conformal boundary ($d=2$). In fact this idea of locality constraints has already been applied in \cite{Lauria:2020emq,Herzog:2022jqv,Behan:2020nsf}. In that case, the defect plays the role of the conformal boundary of AdS$_2$. With a fixed conformal line defect, the (connected part of the) full conformal symmetry is broken to
\[
SO^+(1,2)\times SO(d-1),
\]
which makes the setup very similar to QFT in AdS$_2$: the group $SO^+(1,2)$ acts as the spacetime symmetry, while $SO(d-1)$ is the rotational symmetry transverse to the defect line, also regarded as the internal global symmetry along the defect line.

Likewise, one can define the $\rho$ and $\chi$ variables for the bulk--defect--defect correlator.\footnote{Our $\rho$ and $\chi$ variables are related to the $\nu$ ($\nu_E$) and $\zeta$ ($\zeta_E$) variables in \cite{Bartlett-Tisdall:2025iqx} by
\begin{equation*}
    \zeta=-\zeta_E=-\frac{1}{\chi},\qquad
    \nu=i\nu_E=-\, i\frac{\rho}{\sqrt{\chi}}\,.
\end{equation*}}
When the bulk operator is a scalar, the same conformal block expansion, local block expansion and the corresponding locality sum rules then apply.

A slight difference in that setting is that there is no need to assume that the UV bulk theory is a CFT -- it already is, and the bulk dimension $\Delta_{\hat{\Phi}}$ is part of the CFT data rather than UV data.

\subsection{Comments on parity-symmetric theories}
\label{subsec:parity}

The derivation of the locality sum rules \eqref{sumrule:even} and \eqref{sumrule:odd} does not rely on parity symmetry.  
For a general bulk–boundary–boundary correlator, both the even and odd sum rules appear in a nontrivial way: each of them vanishes only after summing over all intermediate states, rather than term by term.

When the theory has parity symmetry, one can classify the bulk and boundary operators into parity-even and parity-odd ones. In this case, the BOE of a parity-even (odd) bulk operator contains only parity-even (odd) boundary operators. On the boundary, the relation between the OPE coefficients is
\be
C_{lij} = (-1)^{\s_l+\s_i+\s_j}C_{lji}\,,
\ee
where $\s_i$ is the parity of operator $\OO_i$. Depending on the parities of the external operators, we have either $C_{ijl}=C_{jil}$ for all $l$, or $C_{ijl}=-C_{jil}$ for all $l$ (whenever $b^{\hat{\Phi}}_l\neq0$). Consequently, one set of the sum rules, either even or odd, trivializes for a given correlator in the parity-symmetric theory. The even sum rule is the same as the direct analytic continuation of the general $d$ version, whereas the odd sum rule is new.

While the sum rules with parity straightforwardly follow from the more generic version \eqref{sumrule:even} and \eqref{sumrule:odd}, the perspective of how locality constraints arise is slightly different. Let us illustrate using the toy model in section \ref{sec:intro}. In \eqref{eq:y expansion of f(x)}, if we further impose that $f(x)$ is an even function in $x$, then only the first sum in \eqref{eq:y expansion of f(x)}, containing even powers of $x$, can appear. For an odd function $f(x)$, only the second sum in \eqref{eq:y expansion of f(x)}, containing odd powers of $x$, can appear. 

Similarly, if we impose parity invariance and consider a bulk-boundary-boundary correlator with even total parity, then in the $\chi\to1$ expansion of $\GG_{ij}^{\hat\F}(\r)$ (defined in \eqref{bkbdbd:generalform}) only even powers of $\sqrt{1-\chi}$ can appear. On the other hand, the 1D conformal block \eqref{def:bkbdbd block} has the following expansion near $\chi=1$
\al{
G_{\D_l}^{\D_i \D_j}(\chi) = 
\sum_{n\in \mathbb{N}} a_n\, \left(\sqrt{1-\chi}\right)^{n}\,,
}
which contains both even and odd powers of $\sqrt{1-\chi}$. Conceptually, the sum rules with parity originate from removing the unwanted odd powers of $\sqrt{1-\chi}$ in the conformal blocks. For parity-odd correlators, instead we want to remove the even powers of $\sqrt{1-\chi}$. The local blocks  constructed in \eqref{localblock:bkbdbd_even} and \eqref{localblock:bkbdbd_odd} indeed respect this structure
\al{\spl{
G_{\D_l}^{\D_i\D_j,(\a)}(\chi)=
\sum_{\substack{n\geqslant0 \\\text{even}}}b_n \, \left(\sqrt{1-\chi}\right)^{n}\, ,
\qquad
\abs{\r}\tilde{G}_{\D_l}^{\D_i\D_j,(\b)}(\chi)=\sum_{\substack{n\geqslant0 \\\text{odd}}} c_n \, \left(\sqrt{1-\chi}\right)^{n}.
}}

% either the parity-even sum rules hold term by term, or the parity-odd sum rules do.  
% Only one of the two sets remains nontrivial for a given correlator in the parity-symmetric theory.

% We call the OPE coefficient $C_{lij}$ parity-odd (even) if the total parity of the operators $\{\mathcal{O}_l,\, \mathcal{O}_i,\, \mathcal{O}_j\}$ is odd (even).

\section{Test of locality sum rules in free theory}
\label{sec:test}

In this section we test the sum rules \eqref{sumrule:even} and \eqref{sumrule:odd} in the theory of a free massive scalar $\hat\f$ in AdS$_2$
\begin{equation}\label{def:freescalar}
	S=\frac{1}{2}\int d^2x \sqrt{g}\left[g^{\mu\nu}\partial_\mu\hat\phi\partial_\nu\hat\phi+m^2\phih^2\right]\,,
\end{equation}
with Dirichlet (D) or Neumann (N) boundary condition. This theory has parity symmetry and $\hat\f$ is parity-even. On the conformal boundary, the only primary operator appearing in the BOE of $\hat{\f}$ is a generalized free field $\phi$, again with even parity. Its scaling dimension $\D_\f$ satisfies
\be
\D_\f(\D_\f-1)=m^2\,,
\quad
\D_\f = 
\begin{cases}
\frac12 + \frac12 \sqrt{1+4m^2}\,, & (\text{D})\,,
\\
\frac12 - \frac12 \sqrt{1+4m^2}\,, & (\text{N})\,.
\end{cases}
\ee
To test the odd sum rule we need parity-odd operators. In the generalized free theory, we need at least the triple-twist family to get parity-odd operators. The double-twist operators $[\phi^2]_p$ with scaling dimension $\D_{[\phi^2]_p}=2\D_\f+2p$ all have even parity. The triple-twist operators $[\phi^3]_q$ with scaling dimension $\D_{[\phi^3]_q}=3\D_\f+q$ have even parity for even $q$ and odd parity for odd $q$.\footnote{In general the triple trace operators $[\f^3]_{n_1,n_2}$ have two quantum numbers as there can be more than one way to distribute the derivatives to obtain a primary operator. Their dimension is $\Delta_{[\f^3]_{n_1,n_2}} = 3\Delta_\f + n_1+n_2$. For $q=n_1+n_2<6$, there is a unique triple trace primary at each level $q$. For $q\geqslant6$, the notation $[\f^3]_q$ actually denotes a collection of operators with degenerate scaling dimensions (see \cite{Loparco:2026fki}). In this paper we only consider $[\f^3]_3$, which is unique.}

For later convenience, let us define the partial sums
\al{
\spl{
&S_e(l_{\rm max})\assign \sum_{l=0}^{l_{\rm max}} b^{\hat\f^2}_l C_{lij} \q^{(\a)}_n(\D_l,\D_{ij},\D_{ji})\,.
\\
&S_o(l_{\rm max})\assign \sum_{l=0}^{l_{\rm max}} b^{\hat\f^2}_l C_{lij} \q^{(\a)}_n(\D_l,\D_{ij}+1,\D_{ji}+1)\,.
}
\label{eq:partial sum rule}}

\begin{figure}[!t]
    \centering
    \includegraphics[width=0.5\linewidth]{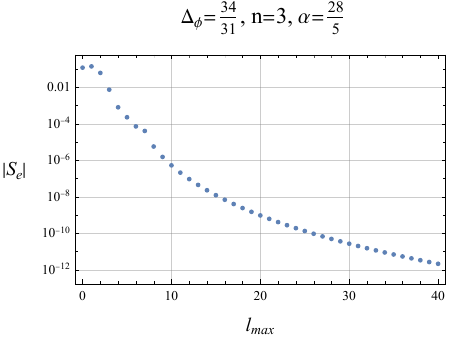}
    \caption{Numerical test of the even sum rule \eqref{sumrule:even}. Here $S_e(l_{\rm max})$ is the partial sum given in \eqref{eq:partial sum rule}.}
    \label{fig:even sum rule test}
\end{figure}

\paragraph{Even sum rule test}
The simplest example to test the even sum rule \eqref{sumrule:even} is the correlator $\<\hat\f^2(\t,z) \f(\t_1) \f(\t_2)\>$. It has the conformal block decomposition
\be
\<\hat\f^2(\t,z) \f(\t_1) \f(\t_2)\>
=
\frac{1}{|\t_{12}|^{2\D_\f}}
\sum_{l\geqslant0} b^{\hat\f^2}_{[\f^2]_l} C_{[\f^2]_l \f \f} G^{\D_\f \D_\f}_{2\D_\f+2l}(\chi)\,.
\ee
Both the BOE coefficient and the OPE coefficient are known explicitly. They are \cite{Meineri:2023mps} 
\begin{equation}\label{eq:BOEphi2}
b^{\hat\phi^2}_{[\f^2]_n}
    =
    \frac{\Gamma(\Delta_\f)}{\Gamma\left(\Delta_\f+\frac{1}{2}\right)}\,
    \sqrt{\frac{\left(\frac{1}{2}\right)_n\left(\Delta_\f\right)_n^3\left(2\Delta_\f-\frac{1}{2}\right)_n}{2\pi\,n!\,\left(2\Delta_\f\right)_n\,\left(\Delta+\frac{1}{2}\right)_n\,\left(\Delta_\f+\frac{1}{4}\right)_n\,\left(\Delta_\f-\frac{1}{4}\right)_n}}\,.
\end{equation}
and \cite{Fitzpatrick:2011dm}
\begin{align}
C_{\f\f[\f^2]_n}
  &= (-1)^n\sqrt{\frac{(2\Delta_\f)_n(2\Delta_\f)_{2n}}{2^{2n-1}(2n)!(2\Delta_\f+n-1/2)_{n}}}\,.
\end{align}
The numerical result is shown in figure \ref{fig:even sum rule test} and \eqref{sumrule:even} is satisfied as expected.

\paragraph{Odd sum rule test}
To test the odd sum rule \eqref{sumrule:odd}, the simplest parity-odd correlator one can consider is $\<\hat\f^2(\t,z) [\f^3]_3(\t_1) \f(\t_2)\>$. However, in this case the odd sum rule is trivially satisfied. The conformal block decomposition of this correlator contains only two primaries: $[\phi^2]_0$ (with dimension $2\Delta_\phi$) and $[\phi^2]_1$ (with dimension $2\Delta_\phi+2$), because the OPE coefficient $C_{[\f^2]_l[\f^3]_q\f}$ is only non-zero for $2l\leq q$. One can check that $\theta^{(\alpha)}_n$'s for the odd sum rules contain a factor of $\G(\frac{1+2l-q}{2})^{-1}$ and all vanish for these cases. Furthermore, the $\chi\to1$ expansion of each of the two blocks only contains odd powers of $\sqrt{1-\chi}$, so they respect parity individually. This type of correlators with a finite number of conformal blocks in its expansion are closely analogous to the free \emph{massless} scalar theory case considered in \cite{Lauria:2020emq,Behan:2020nsf}, where the equation of motion imposes that the BOE of the bulk scalar can at most contain two primaries. In this type of cases, the sum rules need to trivialize because replacing the blocks by local blocks in \eqref{localblock:rewrite} now adds a finite sum to the full correlator, which does not vanish universally for different sets of OPE and BOE coefficients unless each $\q_n^{(\a)}$ is zero. Nevertheless, the locality constraints still exist, and they are embodied either in the relation among the block expansion coefficients, or the spectrum of the block expansion.

Next we consider the three-point function $\<\hat\f^2(\t,z) [\f^3]_3(\t_1) \f^3(\t_2)\>$.
% \footnote{Notice that the simpler correlator $\<\hat\f^2(\t,z) [\f^3]_3(\t_1) \f(\t_2)\>$ trivially satisfies the odd sum rule. The conformal block decomposition of this correlator contains only two primaries: $[\phi^2]_0$ (with dimension $2\Delta_\phi$) and $[\phi^2]_1$ (with dimension $2\Delta_\phi+2$), because the OPE coefficient $C_{[\f^2]_l[\f^3]_q\f}$ is only non-zero for $2l\leq q$. One can check that $\theta^{(\alpha)}_n$'s for the odd sum rules contain a factor of $\G(\frac{1+2l-q}{2})^{-1}$ and all vanish for these cases. {\color{blue} This type of correlators with a finite number of conformal blocks in its expansion are closely analogous to the free (massless) scalar theory case considered in \cite{Lauria:2020emq,Behan:2020nsf}. In this case, the sum rules need to trivialize because replacing the blocks by local blocks in \eqref{localblock:rewrite} now adds a finite sum to the full correlator, which does not vanish universally for different sets of OPE and BOE coefficients unless each $\q$ is zero. Instead, the locality constraints are embodied either in the relation among the block expansion coefficients, or the spectrum of the block expansion, or both.}}
For $\r<0$, it admits the conformal block decomposition
\be
\<\hat\f^2(\t,z) [\f^3]_3(\t_1) \f^3(\t_2)\>
=
\frac{1}{|\t_{12}|^{6\D_\f+3}}
\left(\frac{(\t-\t_2)^2+z^2}{(\t-\t_1)^2+z^2}\right)^{\frac{3}{2}}
\sum_{l=0}^\infty b^{\hat\f^2}_{[\f^2]_l} C_{[\f^2]_l [\f^3]_3 \f^3} G^{3\D_\f+3,3\D_\f}_{2\D_\f+2l}(\chi)\,,
\label{eq:block decomp of free theory correlator}
\ee
which has infinitely many terms. Notice that $C_{[\f^2]_l [\f^3]_3 \f^3} = - C_{[\f^2]_l \f^3 [\f^3]_3 }$.

To test the sum rule, we first need to extract the OPE coefficient $C_{[\f^2]_l [\f^3]_3 \f^3}$ from the block expansion \eqref{eq:block decomp of free theory correlator}. The triple twist operator has the explicit form \cite{Antunes:2025iaw}
\al{\spl{
[\f^3]_3&=
\mathcal{N}_{[\f^3]_3}\left( \phi^2\partial^3\phi-\frac{3(1+\Delta_\f)}{\Delta_\f}\phi\partial\phi\partial^2\phi+\frac{(1+\Delta_\f)(1+2\Delta_\f)}{\Delta_\f^2}(\partial\phi)^3\right) \,,
}
\label{eq:[phi^3]_3}}
with\footnote{We have chosen the normalization convention
\be
\<\OO(x)\OO(y)\> = \frac{1}{|x-y|^{2\D_\OO}}\,,
\ee
so that the OPE coefficient is purely real for parity-even case and purely imaginary for parity-odd case. This is consistent with our convention for operator conjugation in equal-time quantization \eqref{eq:operator conjugation in equal time}. From unitarity, the three-point functions on the boundary satisfy (assuming $\t_1<\t_2<\t_3$)
\al{\spl{
\<\OO_i(\t_1)\OO_j(\t_2)\OO_k(\t_3)\>^\dagger
=\<\OO_k(-\t_3)\OO_j(-\t_2)\OO_i(-\t_1)\>
\ \Rightarrow\ 
C_{ijk}^*= C_{kji}\,.
}}
}
\begin{equation}
    \mathcal{N}_{[\f^3]_3}=2i \sqrt{\frac{6 (\Delta_\phi +1) (2 \Delta_\phi +1) (3 \Delta_\phi +1) (3 \Delta_\phi +2)}{\Delta_\phi }}\, .
\end{equation}
Using \eqref{eq:[phi^3]_3}, the correlator can be computed using Wick contractions 
\begin{equation}
\<\hat\f^2(\t,z) [\f^3]_3(\t_1) \f^3(\t_2)\>
=
-\frac{2 \sqrt{\frac{6}{\pi }} \Gamma (\D_\f +2) }{\NN_{[\f^3]_3} \Gamma \left(\D_\f +\frac{1}{2}\right)}
(\chi -1)^2 
\chi ^{\D_\f } \left(2 \D_\f  (\chi -1)+4 \chi -1\right)
\,,
\end{equation}
where we have put the operators at
\begin{equation}
(\t,z)=\left(0,\sqrt{\frac{\chi}{1-\chi}}\,\right)\,,\qquad \t_1 = 1\,,\qquad \t_2 =\infty\,.
\end{equation}

\begin{figure}[!t]
  \centering
  \begin{subfigure}[t]{0.49\textwidth}
    \includegraphics[width=\textwidth]{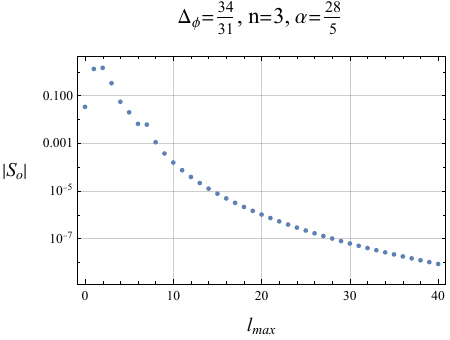}
    \caption{}
  \end{subfigure}
  \hfill
  \begin{subfigure}[t]{0.49\textwidth}
    \includegraphics[width=\textwidth]{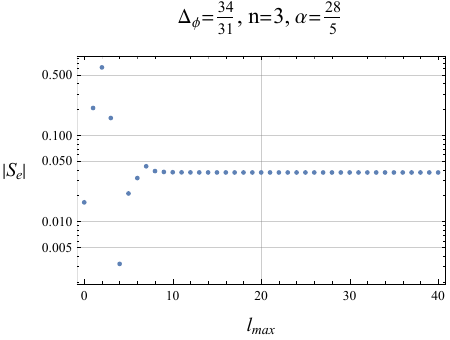}
    \caption{}
  \end{subfigure}
  \caption{$(a)$ Numerical test of the sum rule \eqref{sumrule:odd}.  $(b)$ Illustration that the analytic continuation to $d=1$ of the general $d$ sum rule \eqref{eq:locality sum rule} does not work for the cases where $C_{lij}=-C_{lji}$. Here $S_e(l_{\rm max})$ and $S_o(l_{\rm max})$ are partial sums of the even and odd sum rules, see \eqref{eq:partial sum rule}.}
  \label{fig:odd sum rule test}
\end{figure}

By expanding in small $\chi$ and dividing out \eqref{eq:BOEphi2} we find
\al{\spl{
C_{[\f^2]_l [\f^3]_3 \f^3}
&=
2i  \D_\f  (-1)^l \,
\sqrt{ \frac{\left(\D_\f +1\right)_{l+1} \left(\D_\f +\frac{1}{2}\right)_{l+1}}{\left(2 \D_\f + l-\frac{1}{2}\right)_l} }
\times
\\
&\quad\times
\sqrt{\frac{(\D_\f +l) (\D_\f +l+1) (2 \D_\f +2 l+1)    \Gamma (2 \D_\f +l)}{(\D_\f +1) (2 \D_\f +1) (3 \D_\f +1) (3 \D_\f +2) (2 l)! \Gamma (2 \D_\f ) }}\,.
}}
The numerical test of the sum rule \eqref{sumrule:odd} is shown in figure \ref{fig:odd sum rule test} $(a)$ and it is indeed satisfied. For comparison, we also test and show in figure \ref{fig:odd sum rule test} $(b)$ that the naive analytic continuation in spacetime dimension $d$ of the general $d$ sum rule \eqref{eq:locality sum rule}, equivalent to using $C_{lij}$ instead of $\frac{C_{lij}+C_{lji}}{2}$ in \eqref{sumrule:even}, does not work.

In this section, for simplicity we only test the sum rules with a parity-symmetric theory. We are content with verifying that the (parity) odd sum rule, which does not have a higher dimensional counterpart, indeed works.

It would be interesting to study our sum rules in models that explicitly break parity in AdS$_2$. One example is the two-scalar theory (with $\varphi^a$ and $a\in \{1,2\}$), where parity-breaking terms can be added. In particular, in this case, the action could contain terms as 
\begin{equation}
    S_\text{parity breaking} \supset \int d^2x \sqrt{g} \,\left(\epsilon_{\mu\nu}\epsilon_{ab}\partial^\mu \varphi^a\partial^\nu \varphi^b\right)\left(\varphi^c\varphi^d\delta_{cd}\right)^{n}\, ,
\end{equation}
where we need $n>0$ for this not to be a total derivative.
%which either breaks unitarity or vanishes at the boundary. 
Another example worth studying is a Dirac fermion $\psi(x)$ with non-trivial mass terms. In this case, the action is given by 
\begin{equation}
    S_f = \int d^2x\sqrt{g}\,\Big[\bar{\psi}(\gamma^\mu \mathcal{D}_\mu )\psi+m_s\bar{\psi}\psi+im_a\bar{\psi}\gamma^5\psi\Big]\, ,
\end{equation}
with the curved space gamma matrices $\gamma^\mu=e^\mu_{\,a}\gamma^a$ and the spinor covariant derivative $\mathcal{D}_\mu = \partial_\mu+\frac{1}{4}\omega_{\mu ab}\gamma^{ab}$ with $\gamma^{ab} =\frac{1}{2}[\gamma^a,\gamma^b]$. Here $e^\mu_{\,a}$ is a vielbein and $\omega_{\mu ab}$ the spin connection.
The mass term proportional to $m_a$ breaks parity.\footnote{It can be removed with a chiral rotation of the fermion, but generically this will change the boundary conditions.}

\section{Conclusion and discussion}
\label{sec:conclusion}

In this paper we point out a subtlety in the AdS$_{d+1}$ locality sum rules when restricting to $d=1$ case. On the 1d conformal boundary, the OPE coefficients depend on the operator ordering modulo cyclic permutations, thus those appearing in the locality sum rule \eqref{eq:locality sum rule} must be handled with care. Furthermore, the conformal blocks of bulk-boundary-boundary correlators do not have unphysical singularities to remove, yet the usual conformal block decomposition still diverges when the bulk operator touches the geodesic connecting the two boundary points. Splitting the correlator into an even and an odd part, we derive a dispersion relation \eqref{G:dispersion} for each of them using the power-law bound proved in section \ref{subsec:uniformbound}. The dispersion relation is swappable with the conformal block decomposition due to the power-law bound and the absolute convergence of the BOE, thus leading to a new expansion of the correlator into even and odd local blocks \eqref{localblock:bkbdbd_even} and \eqref{localblock:bkbdbd_odd}. We then derive the even and odd sum rules \eqref{sumrule:even} and \eqref{sumrule:odd} by the fact that the normal and local block expansions must agree. We test the sum rules in the free scalar theory in section \ref{sec:test}. 

The sum rules can be used to derive an integration kernel, which constructs the local blocks through an integral transform of the normal blocks. One concrete example is the local blocks of bulk-boundary-boundary correlators in \eqref{localblock:int_rep}. More generally, we can use the sum rules to improve the convergence of a sum of the form
\begin{align}
\spl{
\sum\limits_{l}\bb{\hat{\Phi}}{l} C_{ijl} F(\D_l) &=
\sum\limits_{l}\bb{\hat{\Phi}}{l}\frac{C_{ijl}+C_{jil}}{2}
  F(\D_l) +
\sum\limits_{l}\bb{\hat{\Phi}}{l}
\frac{C_{ijl}-C_{jil}}{2}F(\D_l)\\
&=\sum\limits_{l}\bb{\hat{\Phi}}{l}\frac{C_{ijl}+C_{jil}}{2}\left[
  F(\D_l)
  -\sum_{n=0}^\infty \theta^{(\a)}_n(\D_l;\D_{ij},\D_{ji}) F(2\a+2n)  
  \right]\\
  &\qquad +\sum\limits_{l}\bb{\hat{\Phi}}{l}
\frac{C_{ijl}-C_{jil}}{2}
\left[
  F(\D_l)
  -\sum_{n=0}^\infty \theta^{(\a)}_n(\D_l;\D_{ij}+1,\D_{ji}+1) F(2\a+2n)  
  \right]\\
  &=\sum\limits_{l}\bb{\hat{\Phi}}{l}\frac{C_{ijl}+C_{jil}}{2}
  \int_C\frac{d\D'}{2\pi i} K^{(\a)}(\D',\D_l;\D_{ij},\D_{ji})
  F(\D')\\
&\qquad +\sum\limits_{l}\bb{\hat{\Phi}}{l}
\frac{C_{ijl}-C_{jil}}{2}
\int_C\frac{d\D'}{2\pi i} K^{(\a)}(\D',\D_l;\D_{ij}+1,\D_{ji}+1)
  F(\D')\,.
}
\end{align}
where in the last line, the contour surrounds the poles in a way similar to figure \ref{fig:integral transform}.

For example, consider a bulk-bulk-boundary correlator $\langle \hat{\Phi}(\tau_1,z_1) \hat{\mathcal{O}}(\t_2,z_2)\mathcal{O}_i(\t_3) \rangle$. Using the BOE for both $\hat\Phi$ and $\hat\OO$ we obtain, similar to \eqref{bkbdbd:exp}, the following conformal block decomposition
\al{
\spl{
    &\langle \hat{\Phi}(\tau_1,z_1) \hat{\mathcal{O}}(\t_2,z_2)\mathcal{O}_i(\t_3) \rangle
    =\begin{cases} 
    \sum\limits_{j,l} \bb{\hat{\Phi}}{l}\bb{\hat{\mathcal{O}}}{j} C_{lji} 
    R_{\D_l,\D_j}^{\D_i}(\t_1,z_1,\t_2,z_2,\t_3)\,, & \upsilon<0\,,
    \\[1em]
    \sum\limits_{j,l} \bb{\hat{\Phi}}{l}\bb{\hat{\mathcal{O}}}{j} C_{lij} 
    R_{\D_l,\D_j}^{\D_i}(\t_1,z_1,\t_2,z_2,\t_3)\,, & \upsilon>0\,,
    \end{cases}
}}
where $R_{\D_l,\D_j}^{\D_i}$ denotes the bulk-bulk-boundary conformal block whose explicit expression is inessential here. The natural cross ratio of bulk-bulk-boundary correlators is $\upsilon$, and ${\rm sgn}(\upsilon)$ is the analogy of ${\rm sgn}(\r)$ in \eqref{bkbdbd:exp}. For more details, see \cite{Loparco:2026fki}. Applying the general formula above to the sum over $l$ we get (suppressing the spacetime coordinates for simplicity)
\begin{align}
    &\langle \hat{\Phi}(\tau_1,z_1) \hat{\mathcal{O}}(\t_2,z_2)\mathcal{O}_i(\t_3) \rangle \\
  &\quad   =\sum\limits_{l,j}  \bb{\hat{\Phi}}{l}\,\bb{\hat{\mathcal{O}}}{j} \Bigg( 
    \frac{C_{lji}+C_{lij}}{2}
  \int\frac{d\D'}{2\pi i} K^{(\a_{ij})}(\D',\D_l;\D_{ij},\D_{ji})
 R_{\D',\D_j}^{\D_i}\nonumber\\
&\quad\phantom{=}+
\frac{C_{lji}-C_{lij}}{2} {\rm sgn}(\upsilon)
\int\frac{d\D'}{2\pi i} K^{(\a_{ij})}(\D',\D_l;\D_{ij}+1,\D_{ji}+1)
 R_{\D',\D_j}^{\D_i}\Bigg)\,.\nonumber
\end{align} 
This is a local block expansion of the bulk-bulk-boundary correlator. It should be possible to also improve the convergence of the sum over $j$ using another kernel. The same idea can also be applied to other correlators, both higher-point and in higher spacetime dimensions. We leave these for future work. 

The sum rules \eqref{sumrule:even} and \eqref{sumrule:odd} can be used to set up a bootstrap program, either for local QFTs in AdS$_2$ or for boundary CFTs (BCFTs) in 2d. While there is no positivity in the sum rules, it is still possible to determine the ratios of  some data (BOE and OPE coefficients) when the BOE spectrum is of generalized free theory type (having even integer gaps), leading to truncation in the sum rules. See \cite[section 5 \& 6]{Levine:2023ywq} for examples. Alternatively, to leverage constraints from positivity one can set up a mixed correlator bootstrap problem \cite[section 7]{Meineri:2023mps} that is the QFT-in-AdS analogy of the form factor S-matrix bootstrap \cite{Karateev:2019ymz}. In particular, the bulk-boundary-boundary three-point functions are analogous to flat-space form factors.

Our original motivation for constructing these sum rules, however, more closely parallels the S-matrix bootstrap program of the 1960s: instead of trying to carve out the space of allowed theories, we aim to solve one specific family of theories at a time. In \cite{Loparco:2026fki}, we study the RG flows of QFTs in AdS$_2$ induced by a relevant bulk operator. We derive a set of universal ordinary differential equations (ODEs) that non-perturbatively encode the evolution along the flow of all the QFT data $\{\D_i,C_{ijk},b^{\hat\OO}_i\}$: the scaling dimensions of boundary operators $\D_i$, the boundary OPE coefficients $C_{ijk}$ and the BOE coefficients $b^{\hat\OO}_i$. There we need to deal with integrals of various correlators over AdS$_2$. The sum rules and local blocks developed in this paper are useful for obtaining a convergent final result.

\section*{Acknowledgments}
We would like to thank Edoardo Lauria, Marco Meineri, Miguel Paulos, and Mingchen Xia for useful discussions. We especially thank Balt van Rees for pointing us to the sector version of the Phragmén–Lindelöf theorem. We also thank the participants of the workshop QFT in AdS 2024 in Trieste, as well as the participants of the Bootstrap 2025 conference in São Paulo, for stimulating discussions. GM and JQ thank Riken iTHEMS and the Yukawa Institute for Theoretical Physics at Kyoto University. Discussions during “Progress of Theoretical Bootstrap” were useful in completing this work.

This work was performed in part at Aspen Center for Physics, which is supported by National Science Foundation grant PHY-2210452 and by a grant from the Simons Foundation (1161654, Troyer).

ML, GM and JP are supported by the Simons Foundation grant 488649 (Simons Collaboration on the Nonperturbative Bootstrap) and the Swiss National Science Foundation through the project
200020\_197160 and through the National Centre of Competence in Research SwissMAP.
The research of ML was also supported by the Italian Ministry of University and Research (MUR) under the FIS grant BootBeyond (CUP: D53C24005470001) and by the
INFN “Iniziativa Specifica” ST\&FI. The work of JQ was supported by World Premier International Research Center Initiative (WPI), MEXT, Japan. JQ also acknowledges support by Simons Foundation grant 994310
(Simons Collaboration on Confinement and QCD Strings), under which a portion of this work was performed. The work of XZ is supported by an ANR grant from the Tremplin - ERC Starting Grant funding scheme.

\appendix

\section{Phragmén–Lindelöf theorem (sector version)}\label{app:PL}

This appendix provides an introduction to the sector version of the Phragmén–Lindelöf theorem (theorem~\ref{theorem:PL}), following \cite[section~5.6]{titchmarsh1939theory}.

Let $f(z)$ be an analytic function on the sector domain \eqref{def:sector}, continuous up to its boundary.  
Without loss of generality, we take $\theta_{1}=-\kappa$ and $\theta_{2}=\kappa$.

Consider the function
\begin{equation}
	\begin{split}
		F_{\epsilon,\gamma}(z):=e^{-\epsilon z^{\gamma}}f(z),\quad \text{with }\epsilon>0\, ,\ \beta<\gamma<\frac{\pi}{2\kappa}\,.
	\end{split}
\end{equation}
By construction,\footnote{Here we consider $\beta\geqslant0$ because the theorem is trivial for $\beta<0$.} 
\begin{equation}
	\begin{split}
		\abs{F_{\epsilon,\gamma}(z)}\leqslant e^{-\epsilon\abs{z}^{\gamma}\cos(\gamma\kappa)}\abs{f(z)}\,,\qquad(\abs{\arg(z)}\leqslant\kappa)\,.
	\end{split}
\end{equation}
Then, by the assumptions of theorem~\ref{theorem:PL},
% \JQ{$<$ vs $\leqslant$ makes no difference here because the function is continuous up to the boundary}
\begin{equation}
	\begin{split}
		 \abs{F_{\epsilon,\gamma}(z)}&\leqslant A e^{-\epsilon\abs{z}^{\gamma}\cos(\gamma\kappa)+\abs{z}^{\beta}}\, , 
		 \qquad (\abs{\arg(z)}\leqslant \kappa)\,, \\
		 \abs{F_{\epsilon,\gamma}(z)}&\leqslant C \, ,\qquad \qquad \qquad \qquad \hspace{0.45cm}(\arg(z)=\pm\kappa)\,.
	\end{split}
\end{equation}
It follows from the first bound that, for sufficiently large $R$,
% {\color{blue}and appropriately chosen $\e$ and $\g$}
\begin{equation}
	\begin{split}
		\abs{F_{\epsilon,\gamma}(z)}\leqslant C\, ,\qquad\quad (\abs{z}\geqslant R,\ \abs{\arg(z)}\leqslant \kappa)\,.
	\end{split}
\end{equation}
Now we apply the maximum-modulus theorem on the region 
\(\{\abs{z}\leqslant R,\ \abs{\arg(z)}\leqslant\kappa\}\) and obtain
\begin{equation}
	\begin{split}
		\abs{F_{\epsilon,\gamma}(z)}\leqslant C\,,\qquad \text{for }\abs{\arg(z)}\leqslant\kappa\,.
	\end{split}
\end{equation}
Returning to the function $f(z)$, we conclude that
\begin{equation}
	\begin{split}
		\abs{f(z)}\leqslant C e^{\epsilon\abs{z}^{\gamma}},\qquad\quad \text{for }\abs{\arg(z)}\leqslant\kappa\,.
	\end{split}
\end{equation}
Since $\epsilon$ can be chosen arbitrarily small, the above estimate implies $\abs{f(z)}\leqslant C$ throughout the same region. This completes the proof of theorem~\ref{theorem:PL}.

\bibliography{bibliography}
\bibliographystyle{utphys}
\end{document}